\documentclass{emulateapj}
\usepackage{amsmath}

\newcommand{\be}{\begin{displaymath}}
\newcommand{\ee}{\end{displaymath}}

\def\lsim{\hbox{\rlap{\raise 0.425ex\hbox{$<$}}\lower 0.65ex\hbox{$\sim$}}}
\def\gsim{\hbox{\rlap{\raise 0.425ex\hbox{$>$}}\lower 0.65ex\hbox{$\sim$}}}

\def\arcsec{\hbox{$^{\prime\prime}$}}

\shorttitle{Asphericity, Interaction, and Dust in SN\,2013ej}
\shortauthors{Mauerhan et al.}

\begin{document}

\title{Asphericity, Interaction, and Dust in the Type II-P/II-L Supernova 2013ej in Messier\,74}

\author{Jon C. Mauerhan\altaffilmark{1, 2},        %mauerhan@astro.berkeley.edu
Schuyler D. Van Dyk\altaffilmark{3},
Joel Johansson\altaffilmark{4},
Maokai Hu\altaffilmark{5},
Ori D. Fox\altaffilmark{6},
Lifan Wang\altaffilmark{5},
Melissa L. Graham\altaffilmark{1,7},                 %melissalynngraham@gmail.com
Alexei V. Filippenko\altaffilmark{1},           %afilippenko@berkeley.edu
and Isaac Shivvers\altaffilmark{1},                 %ishivvers@berkeley.edu
}

\altaffiltext{1}{Department of Astronomy, University of California, Berkeley, CA 94720-3411, USA}
\altaffiltext{2}{e-mail: mauerhan@astro.berkeley.edu}
\altaffiltext{3}{Infrared Processing and Analysis Center, California Institute of Technology, 1200 E. California Blvd., Pasadena, CA, 91125 USA}
\altaffiltext{4}{Department of Particle Physics and Astrophysics, Weizmann Institute of Science, 234 Herzl St., Rehovot, Israel}
\altaffiltext{5}{Department of Physics, Texas A\&M University, College Station, TX 77843, USA}
\altaffiltext{6}{Space Telescope Science Institute, 3700 San Martin Drive, Baltimore, MD 21218, USA }  
\altaffiltext{7}{Department of Astronomy, University of Washington, Box 351580, U.W., Seattle, WA 98195-1580, USA}

\begin{abstract}
Supernova (SN) 2013ej is a well-studied core-collapse supernova (SN) that stemmed from a directly identified red supergiant (RSG) progenitor in galaxy M74. The source exhibits signs of substantial geometric asphericity, X-rays from persistent interaction with circumstellar material (CSM), thermal emission from warm dust, and a light curve that appears intermediate between supernovae of Types II-P and II-L. The proximity of this source motivates a close inspection of these physical characteristics and their potential interconnection. We present multiepoch spectropolarimetry of SN\,2013ej during the first 107 days and deep optical spectroscopy and ultraviolet through infrared photometry past $\sim800$ days. SN\,2013ej exhibits the strongest and most persistent continuum and line polarization ever observed for a SN of its class during the recombination phase. Modeling indicates that the data are consistent with an oblate ellipsoidal photosphere, viewed nearly edge-on and probably augmented by optical scattering from circumstellar dust. We suggest that interaction with an equatorial distribution of CSM, perhaps the result of binary evolution, is responsible for generating the photospheric asphericity.  Relatedly, our late-time optical imaging and spectroscopy show that asymmetric CSM interaction is ongoing, and the morphology of broad H$\alpha$ emission from shock-excited ejecta provides additional evidence that the geometry of the interaction region is ellipsoidal. Alternatively, a prolate ellipsoidal geometry from an intrinsically bipolar explosion is also a plausible interpretation of the data but would probably require a ballistic jet of radioactive material capable of penetrating the hydrogen envelope early in the recombination phase. Finally, our latest space-based optical imaging confirms that the late interaction-powered light curve dropped below the stellar progenitor level, confirming the RSG star's association with the explosion. 
\end{abstract}


\section{Introduction}
Type II-P supernovae (SNe) are the most common class of core-collapse explosion, marking the deaths of red supergiant (RSG) stars with thick hydrogen envelopes. Direct detection of progenitors indicates that their initial stellar masses lie in the range 8--18\,M$_{\odot}$ (Smartt 2009). The namesake \textit{plateau} light curves of SNe~II-P are produced as the photospheric size of the expanding envelope is balanced by the inwardly propagating wave of recombination, releasing the shock-deposited energy over a period of $\sim100$ days. Once the envelope fully recombines and the photosphere passes into the inner layers, the light curve drops steeply and the SN enters the nebular phase, the decline of which is governed by the emission of radioactive decay energy. However, the luminosity evolution during both the recombination and nebular phases can be augmented by interaction with circumstellar material (CSM) lost by the progenitor leading up to core collapse. 

SN\,2013ej is a bright core-collapse supernova (SN) in the nearby galaxy NGC\,628 (Messier\,74). The stellar progenitor was detected in archival imaging data from the \textit{Hubble Space Telescope (HST)} (Van Dyk et al. 2013), and subsequently inferred to be an M-type supergiant with an estimated initial mass in the range 9.5--15.5\,M$_{\odot}$ (Fraser et al. 2014). The SN exhibited a peak absolute magnitude of $\sim-17.5$, and a nebular-phase decline rate consistent with a small mass of radioactive $^{56}$Ni in the range 0.013--0.023\,M$_{\odot}$ (Dhungana et al. 2016; Yuan et al. 2016). Although SN\,2013ej was initially classified as a SN II-P (Leonard et al. 2013; Valenti et al. 2014), subsequent studies showed that the relatively fast decline rate during the recombination phase ($\sim1.2$ mag in the first 50 days after rising up to the plateau; Valenti et al. 2016) and some of the spectroscopic features appear similar to those of SNe from the II-L class (Bose et al. 2015; Huang et al. 2015; Dhungana et al. 2016; Valenti et al. 2016). The distinction between SN~II-L and II-P light curves is not always obvious (Anderson et al. 2014; Sanders et al. 2015); indeed, large samples have revealed a continuum of light-curve morphologies intermediate between sources classified as SNe II-P and SNe II-L. Thus, a simple separation criterion for the two classes is probably not valid (Valenti et al. 2016). It is generally clear, however, that both SN types usually stem from cool supergiant progenitors. The detailed evolution of SN~II light curves (rise time, peak brightness, and plateau/linear recombination phases) is thought to be a combined function of progenitor envelope mass and radius (possibly reduced by pre-SN mass loss in the case of objects classified as SNe~II-L) and explosion energy (Gall et al. 2015) --- but it is also likely that the observational characterization of the SN~II-P and II-L subclasses could be affected by CSM interaction, which can provide an added source of continuum luminosity, and/or aspherical explosion geometries that are viewed at different angles. 

Aspherical supernova geometries are diagnosable with spectropolarimetric observations (see review by Wang \& Wheeler 2008). For SNe II-P, in particular, most objects that have been well studied with this technique have exhibited only weak linear polarization ($\lesssim0.2$\%) during their recombination phases, indicating that the outer envelopes of these explosions tend to be highly spherical (Leonard et al. 2006; Chornock et al. 2010; but also see Dessart \& Hillier 2011). However, substantial increases in continuum and line polarization have been observed in many SNe II-P at the end of the plateau stage and into the nebular phase. It has been suggested that as the electron-scattering photosphere recedes into the ejecta, the more aspherical inner layers become exposed. Nonuniform absorption of the photosphere by ions can yield enhanced polarization of the associated spectral lines. A commonly invoked origin for explosion asphericities is an asymmetric distribution of radioactive $^{56}$Ni from the core, which could manifest itself in a variety of geometric configurations, from irregular plumes or clumps to axisymmetric lobes, tori, or jets (Chugai et al. 1992, 2006). It is important to note, however, that optical scattering of SN light by an aspherical distribution of circumstellar dust is another potential means of producing net continuum polarization (the light-echo effect; e.g., Wang \& Wheeler 1996). This latter mechanism might be particularly important after the direct light from the recombination phase diminishes and the delayed scattering of that light off of dusty CSM continues.

In addition to spectropolarimetry, late-time spectroscopy can probe for the presence of explosion asphericities as well.  If SN ejecta encounter CSM created by the progenitor's pre-SN mass loss, broad-line emission resulting from the passage of ejecta through the reverse shock can be detected (Chevalier \& Fransson 1994; Mauerhan \& Smith 2012; Milisavljevic et al. 2012). The signature of an aspherical interaction zone, either from aspherical  ejecta or CSM, can become imprinted on the emission-line profiles. However, the causes of emission-line asymmetries are not limited to the geometric configuration of the ionized material; the distribution of newly synthesized or pre-existing circumstellar dust, if present, can result in varying degrees of obscuration for the approaching and receding regions of the outflow and can thus have a substantial effect on the line-profile morphologies as well. 

Leonard et al. (2013) were the first to report strong linear polarization from SN\,2013ej, measured as early as day 7. Although no correction for the effects of interstellar polarization (ISP) were applied, substantial intrinsic asphericity was nonetheless convincingly indicated by the data; the implied asphericity for the outer hydrogen envelope is unusual in the context of previously studied SNe II-P. More recently, Kumar et al. (2016) presented $R$-band imaging polarimetry of only the post-plateau nebular phase, reporting null polarization beginning at $\sim100$ days that climbed to $2.15\pm0.57$\% by day $\sim136$; however, no data during the earlier recombination phase were obtained in that study. In addition to indications of asphericity from polarimetry data, SN\,2013ej exhibited signs of asphericity in its optical flux spectrum during its nebular phase, in the form of an asymmetric red shoulder on the H$\alpha$ profile (Bose et al. 2015). 

SN\,2013ej also exhibited signs of weak CSM interaction in X-ray data. Chakraborti et al. (2016) reported the results of long-term X-ray monitoring observations between days 13 and 145, confirming persistent interaction with CSM from a pre-explosion steady wind having a density and mass-loss rate typical of an RSG ($\dot{M}\approx 3 \times 10^{-6}\,{\rm M}_{\odot}$), which must have been blowing for at least 400\,yr. Thermal dust emission is another documented feature of SN\,2013ej. Tinyanont et al. (2016) present measurements of the SN out to day 438, showing the infrared (IR) decline to be broadly consistent with the behavior of most core-collapse SNe.  SN\,2013ej also exhibited some peculiar ionization characteristics at early phases. In particular, Si\,{\sc ii} emission appeared during the first 2 weeks (first noted by Leonard et al. 2013).

In this paper, we investigate the nature of explosion asphericity, CSM interaction, and thermal dust emission in SN\,2013ej. Multi-epoch spectropolarimetry of SN\,2013ej during the recombination and early nebular phases is given in \S2, and we interpret the data in the context of simple electron-scattering and dust-scattering models. In \S3, we present deep late-time optical spectroscopy and ultraviolet (UV) through optical imaging to investigate the persistence of SN emission and CSM interaction. Additional epochs of late-time mid-IR photometry are given in \S4 and compared with existing data, and we investigate the evolution of thermal emission from dust. In \S5, we discuss potential interpretations of the data and the possible physical interconnections between the asphericity, CSM interaction, and thermal dust emission.

\section{Spectropolarimetry}

\subsection{Kast Observations}

Spectropolarimetry was performed at Lick Observatory using the the Shane 3\,m reflector and Kast spectrograph (Miller et al. 1988; Miller \& Stone 1993). SN\,2013ej was observed on the following dates in 2013: Aug. 4, 8, 12, 30; Sep. 6; Oct. 1, 5, 26, Nov. 2, 8 (UTC is used throughout this manuscript). These dates correspond to epochs between 11.5 and 107.2 days, assuming the explosion occurred on July 23.95 (JD = 2,456,497.45; Valenti et al. 2014). Kast is a dual-beam spectropolarimeter that utilizes a rotatable semiachromatic half-waveplate to modulate the incident polarization and a Wollaston prism in the collimated beam to separate the two orthogonally polarized spectra onto the CCD detector.  Only the red channel of Kast was used for spectropolarimetry; a GG455 order-blocking filter suppressed all second-order light at wavelengths shortward of 9000\,{\AA}. 

Observations were made with the 300\,line\,mm$^{-1}$ grating and a 3{\arcsec}-wide slit, yielding a spectral resolution of $\sim4.6$\,{\AA}\,pixel$^{-1}$ and a full width at half-maximum intensity (FWHM) spectral resolution of $\sim16$\,{\AA} (as determined by the FWHM of calibration-lamp lines). The useful wavelength coverage with this setup is 4600--9000\,{\AA}. The orientation of the slit on the sky was always set to a position angle of $180^{\circ}$ (i.e., aligned north-south). In most cases the source was observed through transit of the meridian, and thus tracked through parallactic angle. However, some observations were performed with the slit at a position angle far enough from the parallactic angle that we experience a small amount of blue light loss. This does not affect the spectropolarimetry, owing to the dual-beam instrumentation, but does slightly affect the individual flux spectra we obtain from the data. Exposures of 900\,s were obtained at each of four waveplate positions ($0\fdg0$, $45\fdg0$, $22\fdg5$, and $67\fdg5$). In most cases, three waveplate sequences were performed. Flatfield and calibration-lamp spectra were obtained immediately after each sequence, without moving the telescope. Data were extracted and calibrated using generic IRAF routines and our own IDL functions. Spectropolarimetric analysis was also performed in IRAF and IDL following the methods described by Miller et al. (1988) and implemented by Leonard et al. (2001). All spectral data of the SN have been shifted to the rest-frame wavelength, assuming a host-galaxy redshift of $z=0.00219$.

For polarimetric calibrations, standard stars were selected from the sample of Schmidt et al. (1992a,b). Unpolarized standard stars HD\,212311, BD$+$32\,3739, and HD\,57702 were observed to verify the low instrumental polarization of the Kast spectrograph. We constrained the average fractional Stokes $Q$ and $U$ values to $<0.05$\%. By observing the unpolarized standard stars through a 100\% polarizing filter, we determined that the polarimetric response is so close to 100\% that no correction was necessary, and we obtained the instrumental polarization position-angle curve, which we used to correct the data. We observed the high-polarization stars HD\,204827, HD\,19820, and BD$+$59\,389 to obtain the zeropoint of the polarization position angle on the sky ($\theta$) and to determine the accuracy of polarimetric measurements, which were generally consistent with previously published values to within $\Delta P<0.05$\% and $\Delta \theta<1^{\circ}$. 

\subsection{Polarimetry Formalism}
Linear polarization is expressed as the quadratic sum of the $Q$ and $U$ Stokes vectors, $P= \sqrt{Q^2 + U^2}$. The position angle on the sky is given by $\theta=(1/2)\,\textrm{tan}^{-1}(U/Q)$, taking into account the quadrant in the $Q$--$U$ plane where the inverse tangent angle is located.  Since $P$ is a positive-definite quantity, it is overestimated in situations where the signal-to-noise ratio (S/N) is low.  It is thus typical to express the ``debiased" (or bias-corrected) form of $P$ as $P_{\rm db}=\sqrt{(Q^2 + U^2) - (\sigma_{Q}^2 + \sigma_{U}^2)}$, where $\sigma_Q$ and $\sigma_U$ are the uncertainties in the $Q$ and $U$ Stokes parameters. Note, however, that at low S/N, $P_{\rm db}$ is also not a reliable function, as it has a peculiar probability distribution (Miller et al. 1988). Thus, for extracting statistically reliable values of polarization within a particular waveband, we have binned the calibrated $Q$ and $U$ Stokes spectra separately over the wavelength range of interest before calculating $P$ and $\theta$. All quoted and tabulated values in this paper were determined in this manner, while some figures present $P_{\rm db}$. For $\theta$, if $(\sigma_{Q}^2 + \sigma_{U}^2) > (Q^2 + U^2)$, then we set a 1$\sigma$ upper limit on $P$ of  $\sqrt{\sigma_{Q}^2 + \sigma_{U}^2}$. In cases where $P/\sigma_{P} < 1.5$, $\theta$ is essentially undetermined and is not graphically displayed. 

\begin{figure}
\includegraphics[width=3.3in]{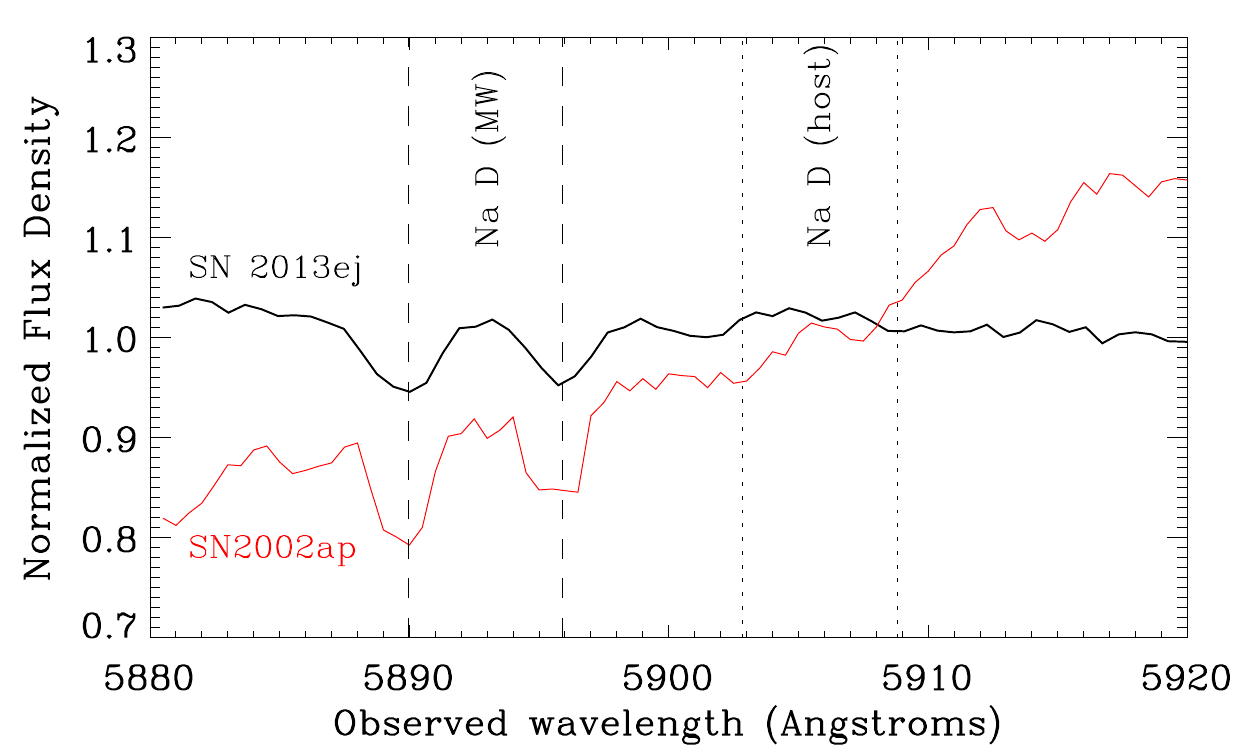}
\caption{Spectral region near Na\,D interstellar absorption feature for SN\,2013ej and SN\,2002ap, both of which occurred in M\,74. \vspace{3mm}}
\label{fig:na}
\end{figure}

 \begin{center}
 \begin{table*} 
      \caption{Integrated Broadband and Continuum Polarization of SN~2013ej.$^{\textrm{a}}$}
\renewcommand\tabcolsep{5.5pt} \scriptsize
\begin{tabular}[b]{@{}llrcrrrrr}
\hline
\hline
UT Date & MJD &Epoch$^{\textrm{b}}$   &  $P_{V}$\,(\%)$^{\textrm{c}}$ & $\theta_{V}$\,(deg)$^{\textrm{c}}$ &  $P_{R}$\,(\%)$^{\textrm{d}}$ & $\theta_{R}$\,(deg)$^{\textrm{d}}$&  $P_{\rm cont}$\,(\%)$^{\textrm{e}}$ & $\theta_{\rm cont}$\,(deg)$^{\textrm{e}}$ \\
\hline
\hline
\multicolumn{9}{|c|}{\bf Observed Values$^f$} \\
\hline
Aug. 04 & 56509.0   & 11.5     & $1.49\pm0.02$ & $96.6\pm0.4$ & $ 1.22\pm 0.02$ &  $97.6\pm 0.3 $& $1.49\pm 0.07$ & $96.6\pm 0.3$ \\  
Aug. 08 & 56513.0   & 15.5     & $1.40\pm0.02$&  $96.9\pm0.6$& $1.12\pm  0.02$ &$ 98.8 \pm 0.4$ &$ 1.40\pm 0.08$ &  $96.9 \pm0.3$ \\
Aug. 12 & 56517.0   &  19.5    & $1.46\pm0.01$&  $96.5\pm0.3$& $ 1.28 \pm 0.01$&$97.8\pm  0.1$  & $1.46\pm 0.03$& $96.5 \pm0.2$ \\
Aug. 30 &  56535.0  &  37.5    & $1.43\pm0.02$&  $96.8\pm0.4$& $1.30 \pm 0.01$&$ 99.2\pm  0.2$ &$ 1.43 \pm0.05$ & $96.8 \pm0.3$  \\
Sep. 6 &  56541.9   &  44.4    & $1.49\pm0.02$&  $97.1\pm0.4$&$1.37\pm  0.01$ &$98.9 \pm 0.2$ &$1.49 \pm0.05$ &$ 97.1 \pm0.2$   \\
Oct. 1 & 56567.0      &  69.5    & $1.68\pm0.03$&  $94.6\pm0.6$& $1.53\pm  0.02$& $ 95.1\pm  0.3$ &$1.68\pm 0.08$ &$94.6\pm 0.4$  \\
Oct. 5 & 56570.8      &  73.3    & $1.44\pm0.03$&  $97.0\pm0.6$& $1.44 \pm 0.02$& $ 98.5\pm  0.2$ &  $ 1.44 \pm0.07$&$97.0\pm 0.4$  \\
Oct. 26 & 56591.8    & 94.3     & $1.51\pm0.02$&  $99.6\pm0.5$& $1.61 \pm 0.01$& $ 102.7 \pm 0.2$ & $1.51\pm 0.05$& $ 99.6 \pm0.3$ \\
Nov. 2 & 56598.8     & 101.3  & $1.51\pm0.03$&  $97.4\pm0.6$& $1.41 \pm 0.02$& $ 97.6 \pm 0.2$&$1.51 \pm0.07$ &  $97.4 \pm0.4$ \\
Nov. 8 & 56604.7     & 107.2  & $1.43\pm0.05$&  $96.0\pm1.1$& $1.35\pm  0.02$& $ 95.7 \pm 0.3$& $1.43\pm0.11$& $96.0 \pm0.7$ \\
\hline
\hline
\multicolumn{9}{|c|}{\bf ISP-Subtracted Values$^f$} \\
\hline
Aug. 04 & 56509.0   & 11.5     & $1.28\pm0.02$ & $86.9\pm0.4$ & $ 1.02\pm 0.02$ &  $85.9\pm 0.5 $& $0.96\pm 0.07$ & $88.0\pm 1.6$ \\
Aug. 08 & 56513.0   & 15.5     & $1.20\pm0.02$&  $86.5\pm0.6$& $0.91\pm  0.02$ &$ 86.0 \pm 0.7$ &$ 0.84\pm 0.08$ &  $88.1 \pm2.7$ \\
Aug. 12 & 56517.0   &  19.5    & $1.26\pm0.01$&  $86.6\pm0.3$& $ 1.08 \pm 0.01$&$86.8\pm  0.3$  & $0.99\pm 0.03$& $87.9 \pm0.7$ \\
Aug. 30 &  56535.0  &  37.5    & $1.23\pm0.02$&  $86.8\pm0.4$& $1.07 \pm 0.01$&$ 88.6\pm  0.4$ &$ 1.03 \pm0.05$ & $85.6 \pm1.1$  \\
Sep. 6 &  56541.9   &  44.4    & $1.28\pm0.02$&  $87.5\pm0.4$&$1.14\pm  0.01$ &$88.9 \pm 0.3$ &$0.97 \pm0.05$ & $89.8 \pm1.1$   \\
Oct. 1 & 56567.0      &  69.5    & $1.49\pm0.03$&  $86.0\pm0.6$& $1.35\pm  0.02$& $ 85.8\pm  0.4$ &$1.41\pm 0.08$ &$87.4\pm 1.1$  \\
Oct. 5 & 56570.8      &  73.3    & $1.23\pm0.03$&  $87.0\pm0.6$& $1.21 \pm 0.02$& $ 89.0\pm  0.4$ &  $ 1.11 \pm0.07$&$91.2\pm 1.3$  \\
Oct. 26 & 56591.8    & 94.3     & $1.25\pm0.02$&  $90.4\pm0.5$& $1.30 \pm 0.01$& $ 95.0 \pm 0.3$ & $1.48\pm 0.05$& $ 101.1 \pm0.6$ \\
Nov. 2 & 56598.8     & 101.3  & $1.29\pm0.03$&  $88.0\pm0.6$& $1.20 \pm 0.02$& $ 87.7 \pm 0.4$&$1.18 \pm0.07$ &  $88.3 \pm1.3$ \\
Nov. 8 & 56604.7     & 107.2  & $1.24\pm0.05$&  $85.7\pm1.2$& $1.17\pm  0.02$& $ 85.1 \pm 0.6$& $1.01 \pm0.11$& $88.3 \pm2.5$ \\
\hline
\end{tabular}\label{tab:p48} 
\begin{flushleft}
 \scriptsize$^\textrm{a}$Uncertainties are statistical.\\
 \scriptsize$^\textrm{b}$Epoch is given in days since the adopted explosion date (JD~2,456,497.45; Valenti et al. 2014).  \\
 \scriptsize$^\textrm{c}$$V$-band values averaged over the wavelength range 5050--5950\,{\AA}. \\
 \scriptsize$^\textrm{d}$$R$-band values averaged over the wavelength range 5890--7270 \,{\AA}. \\
 \scriptsize$^\textrm{e}$Continuum sample region is 7800--8150\,{\AA}; some weak line contamination might be present. \\
 \scriptsize$^\textrm{f}$ $P_{\textrm{\tiny{ISP}}}=0.51$\% and $\theta_{\textrm{\tiny{ISP}}}=125^{\circ}$ values were adopted from Leonard et al. (2002).  
\end{flushleft}
\vspace{4mm}
\end{table*}
\end{center}

\subsection{Interstellar Polarization (ISP)}
As light traverses interstellar media, non-spherical dust grains aligned with the local magnetic field will preferentially scatter incident photons with electric vectors that are aligned with the grains (Hiltner 1949; Davis 1955). This results in the dichroic extinction of light, which polarizes the transmitted signal. The resulting polarization of an object is thus the vector sum of the interstellar and intrinsic source Stokes parameters. Therefore, to obtain a reliable measurement of the intrinsic polarization of an extragalactic source, the interstellar component of the polarization must be determined and subtracted from the observed Stokes vectors. This is often a very problematic element of polarimetric analysis, as an accurate assessment of the ISP from both the Milky Way (MW) and, in particular, the SN host galaxy, can be very difficult to obtain. Foreground stars in the MW can be used to gauge only the Galactic component of ISP.

Prior to SN\,2013ej, M\,74 was host to SN\,2002ap, for which spectropolarimetry was obtained by Leonard et al. (2002), Kawabata et al. (2002), and Wang et al. (2003). Considering the results of Kawabata et al. (2002) along with their own, Leonard et al. (2002) concluded a MW-dominated ISP value of $Q_{\textrm{\tiny{ISP}}}=-0.17$\% and $U_{\textrm{\tiny{ISP}}}=-0.48$\% ($P_{\textrm{\tiny{ISP}}}=0.51$\%, $\theta_{\textrm{\tiny{ISP}}}=125^{\circ}$) in the direction of SN\,2002ap. Wang et al. (2003) arrived at a similar value of $Q_{\textrm{\tiny{ISP}}}=-0.26$\% and $U_{\textrm{\tiny{ISP}}}=-0.55$\% ($P_{\textrm{\tiny{ISP}}}=0.61$\%, $\theta_{\textrm{\tiny{ISP}}}=121^{\circ}$) but concluded that the host ISP is stronger than the MW. To determine whether the ISP estimates for SN\,2002ap can be reasonably applied to SN\,2013ej, we must first confirm that the ISP is indeed MW dominated. Figure\,\ref{fig:na} compares the spectra of both SNe in the spectral region around the Na\,{\sc i}\,D interstellar absorption lines (spectra reproduced from Foley et al. 2003 and Dhungana et al. 2016). No significant absorption features are detected at the redshift of the host, while the lines from the MW have comparable strengths. Since the extinction is clearly dominated by the MW, one could reasonably conclude that the ISP is probably dominated by the MW as well. We therefore adopt values $P_{\textrm{\tiny{ISP}}}=0.51$\% and $\theta_{\textrm{\tiny{ISP}}}=125^{\circ}$ from Leonard et al. (2002), and applied them to our spectropolarimetric data according to the functional form of Serkowski et al. (1975) and Whittet et al. (1992).

However, in their recent broadband imaging polarimetry study of SN\,2013ej,  Kumar et al. (2016) performed an analysis of foreground ISP-probe stars in the broad vicinity of SN\,2013ej, while also considering the ISP values derived by Leonard et al. (2002), Kawabata et al. (2002), and Wang et al. (2003) for the case of SN\,2002ap. Narrowing the sample to four ISP probe stars within 5${^{\prime}}$ of the SN, and after removing those which exhibited evidence for polarimetric variability, Kumar et al. (2016) derived $P_{\textrm{\tiny{ISP}}}=0.64\%$ with $\theta_{\textrm{\tiny{ISP}}}=108.1^{\circ}$ in the $R$ band for SN\,2013ej. If this value is scaled from the $R$ band to the entire spectrum according to Serkowski et al. (1975), then the resulting $V$-band Stokes parameters are $Q_{\textrm{\tiny{ISP}}}=-0.53$\% and $U_{\textrm{\tiny{ISP}}}=-0.39$\% ($P_{\textrm{\tiny{ISP}}}=0.66\%$). This value of $P_{\textrm{\tiny{ISP}}}$ is only slightly higher than the ISP estimates for SN\,2002ap discussed above, and at a position angle that differs by $\sim15^{\circ}$; but it is also 0.1\% larger than the maximum polarization expected from the limiting relation suggested by Serkowski et al., where $P_{\rm max}=9\times E(B-V)=0.56\%$, assuming $E(B-V)=0.062$\,mag in the direction of M74 (Schlafly \& Finkbeiner 2011). We therefore consider this to be an overestimate, and instead we move forward adopting the ISP values derived by Leonard et al. (2002).

A previous employed method to check the validity of ISP estimates is to assume that the peaks of the strongest emission-line features in the flux spectrum, such as H$\alpha$, are intrinsically unpolarized (e.g., see Tran et al. 1997); this method can be particularly useful when the host-galaxy ISP is also significant (note that this does not appear to be the case for SN\,2013ej). The justification for this assumption is that strong emission lines form in the outer regions of the ejecta, well above the polarized continuum photons emerging from the electron-scattering photosphere. However, this assumption might be inappropriate; although substantial depressions in polarization are indeed normally observed at the wavelengths of broad emission components, radiative transfer calculations by Dessart \& Hillier (2011) have demonstrated that variations in optical depth can result in non-null polarization of the H$\alpha$ emission feature. Furthermore, weak features of He that overlap with broad H$\alpha$ might also contribute polarized flux at these wavelengths (Maund et al. 2007). Moreover, the assumption of unpolarized emission-line peaks assumes that electron scattering is the sole source of the polarized flux. For example, if dust scattering in CSM is an important contributor to the net polarization, then the use of this method will obviously result in erroneous estimate of ISP. As we will demonstrate in the following sections, the ISP-subtracted data indeed show a changing amount of residual polarization remaining at the wavelength of the H$\alpha$ flux peak for all epochs, which could be indicative of CSM dust scattering.

\begin{figure*}
\centering
\includegraphics[width=5.5in]{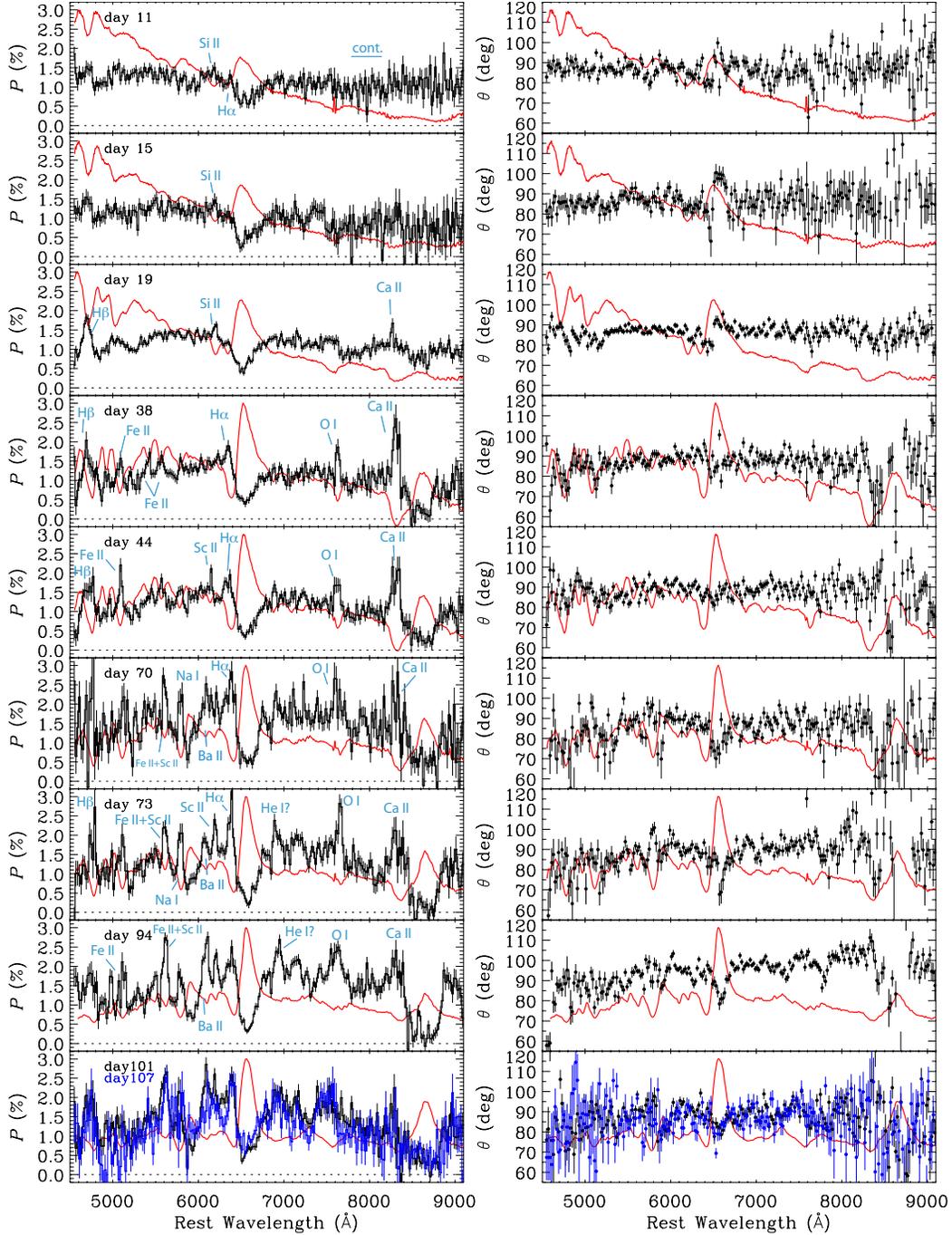}
\caption{Debiased polarization ($P$) and position angle ($\theta$)  for days 11 through 107. The flux spectra are plotted in red for reference. Prominent polarized line features are marked (note that polarized peaks are always blueshifted with respect to the central wavelengths of the associated transitions). Wavelengths for which $P/\sigma_{P}<1.5$ have been omitted from the $\theta$ curve.}
\label{fig:p_theta}
\vspace{3mm}
\end{figure*}

\subsection{Polarization of SN\,2013ej}
The observed and ISP-subtracted values of integrated $P$ and $\theta$ for SN\,2013ej are presented in Table\,1, and the polarized spectra are presented in Figure\,\ref{fig:p_theta} (flux spectra are also included for reference). Hereafter all reference to the polarization data will imply the ISP-subtracted values. Spectral line identifications are adopted from the literature --- in particular, from Bose et al. (2015) and Dhungana et al. (2016), whose line identifications are supported by their spectral modeling. Those line identifications are also corroborated by other spectroscopic studies of SN\,2013ej (e.g., Valenti et al. 2014; Huang et al. 2015; Yuan et a. 2016). We refer the reader to these works for further detailed justification of line identifications.

\begin{figure*}
\includegraphics[scale=0.7]{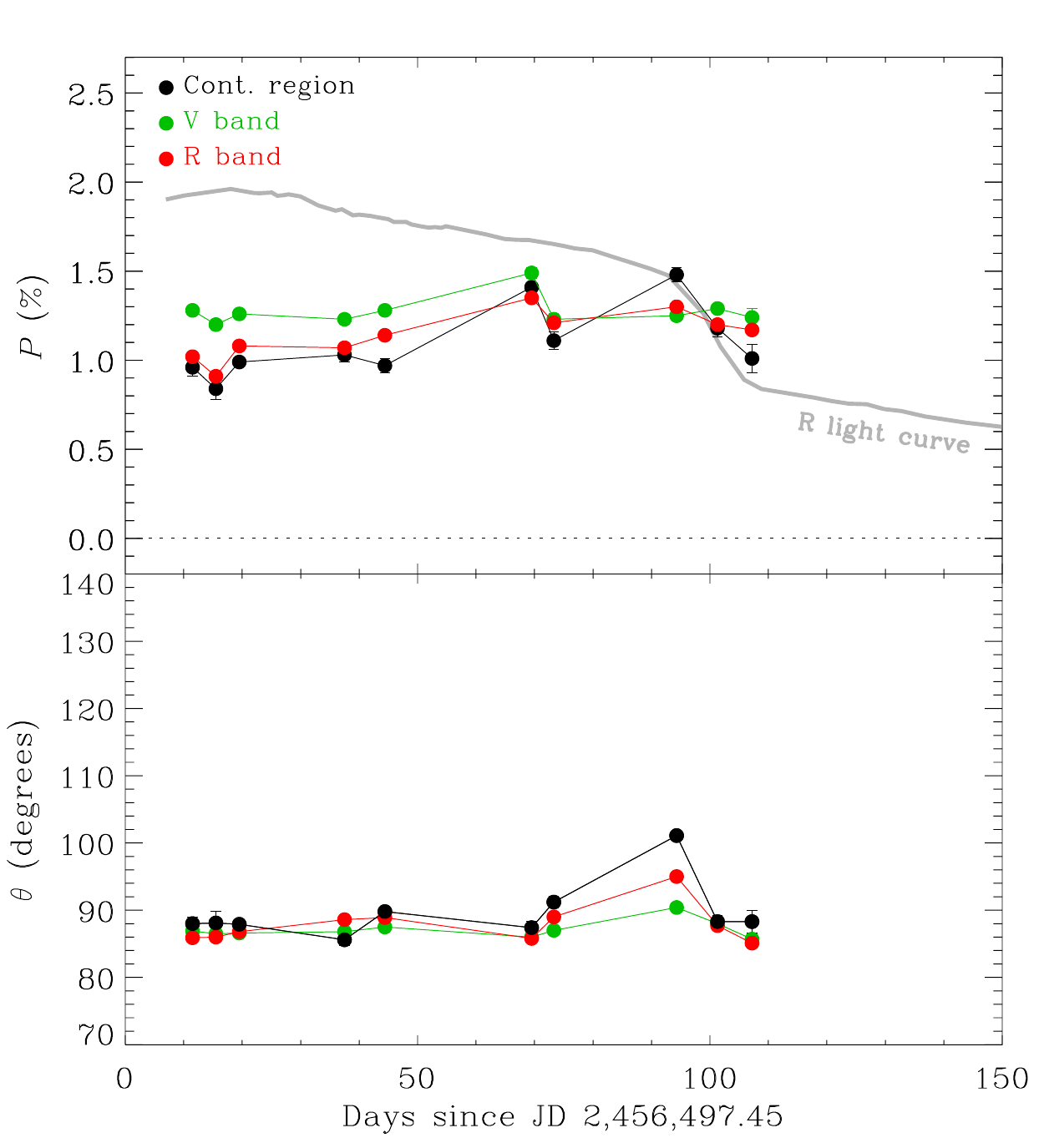}
\includegraphics[scale=0.7]{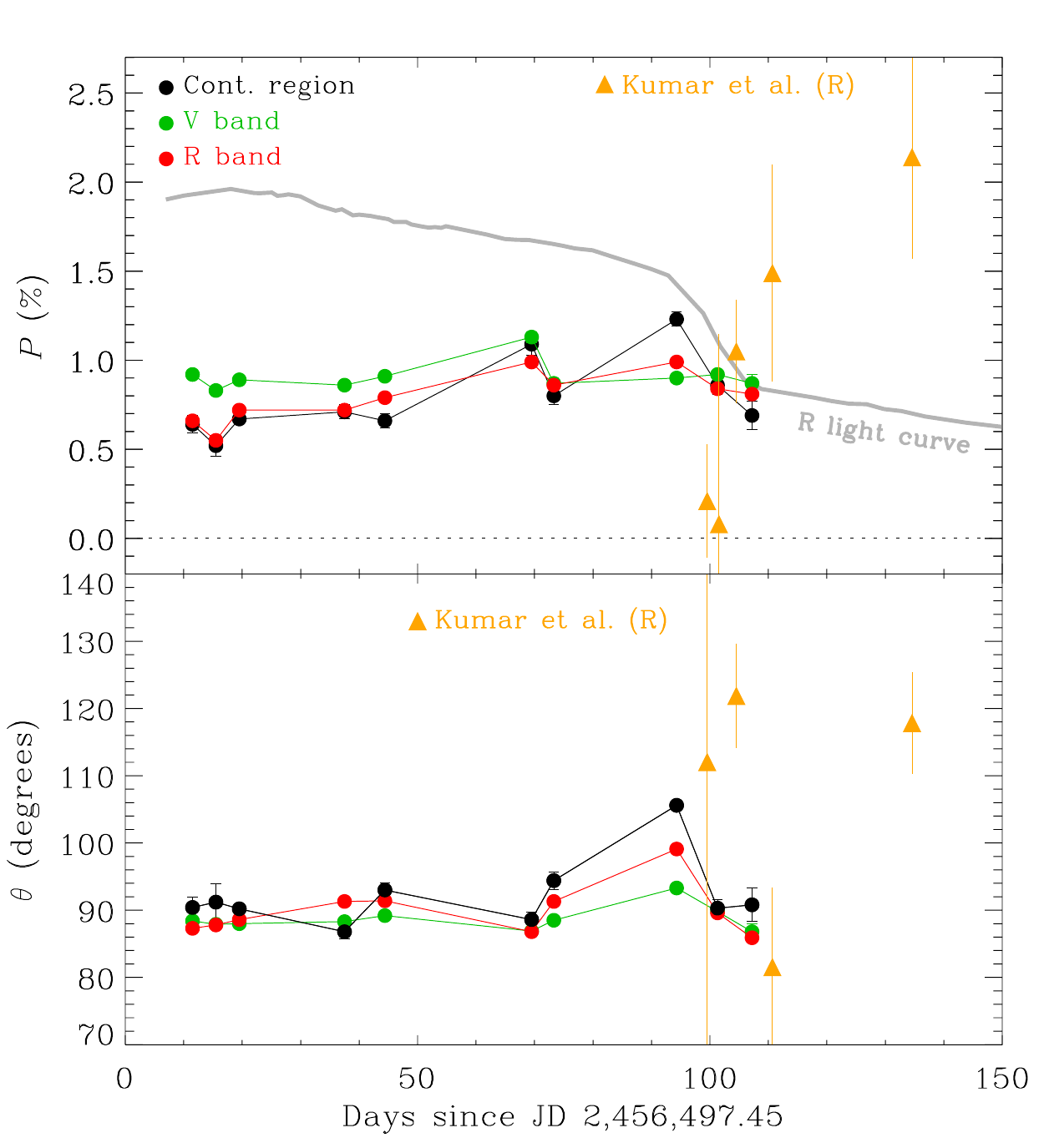}
\caption{\textit{Left}: Temporal evolution of the integrated broadband $P$ and $\theta$ for SN\,2013ej, after subtracting our favored value for the ISP (see \S2.3). The red, green, and black dots represent the R-band, V-band, and continuum regions, respectively. A scaled version of the light curve is shown in gray color for comparison. \textit{Right}: Same data after subtracting the slightly different value of ISP adopted from Kumar et al. (2016), whose $R$-band imaging polarimetry measurements of SN\,2013ej are also shown for comparison (orange triangles).}\vspace{3mm}
\label{fig:pt_lc}
\vspace{3mm}
\end{figure*}

From day 11, we observe substantial polarization of the continuum at a level of $\sim0.9$\%. We chose the wavelength range of 7800--8150\,{\AA} to sample the continuum because this region remains mostly devoid of substantial line features throughout the evolution of the SN. There are modulations in $P$ and $\theta$ across H$\alpha$, H$\beta$, and slightly for He/Na around 5875\,{\AA}. The level of polarization near the peak of the H$\alpha$ emission component of the P-Cygni profile is at $\sim0.5$\%, while $\theta$ rotates by $\sim20^{\circ}$ between the blueshifted and the redshifted sides of the emission feature, straying $\sim20^{\circ}$ from the continuum position angle of the continuum on both sides. The data on day 15 appear similar, but the modulation across H$\alpha$ dips near null levels of $P$ and the rotation across the H$\alpha$ feature increase to $\Delta \theta\approx35^{\circ}$.  

\begin{figure*}
\includegraphics[width=3.5in]{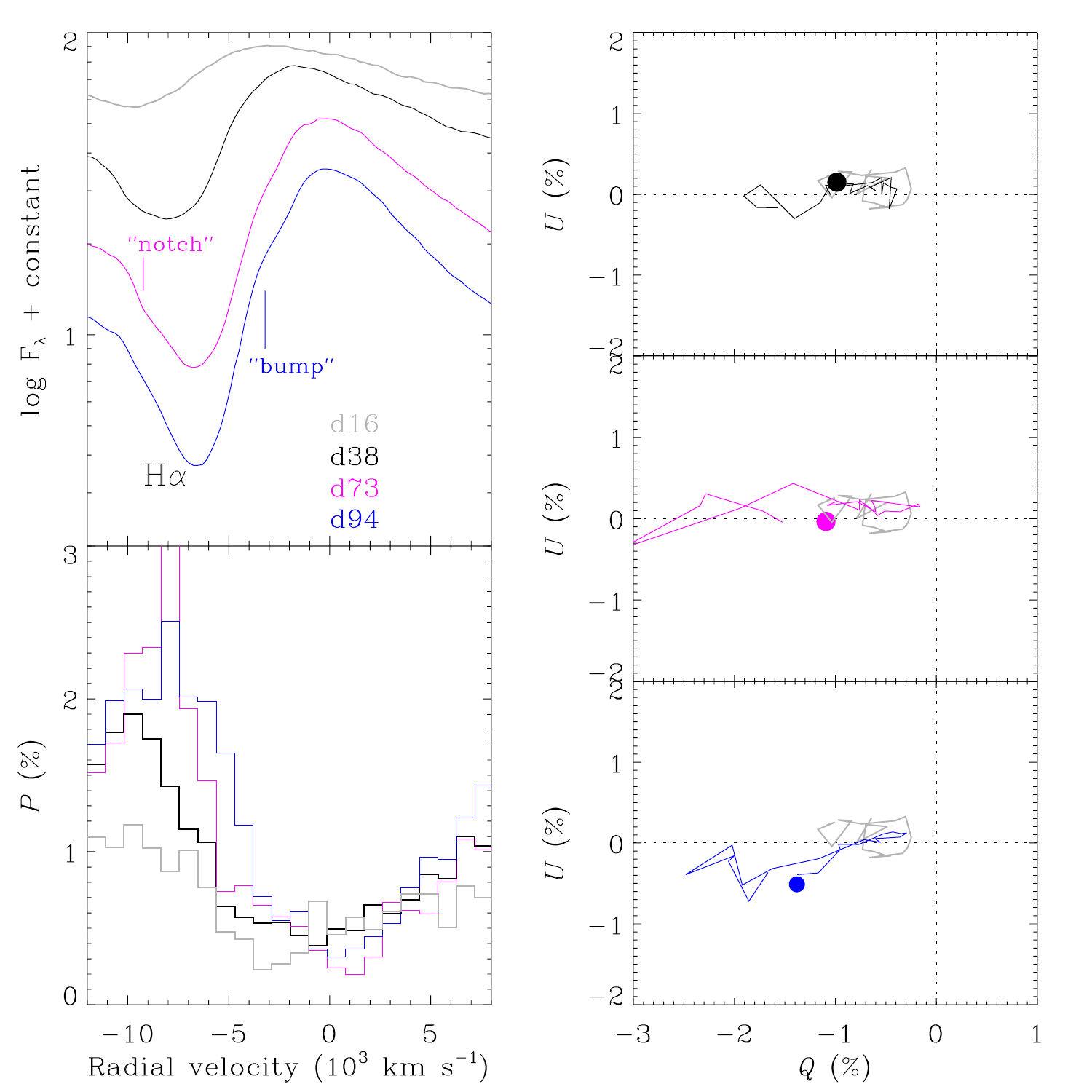}
\includegraphics[width=3.5in]{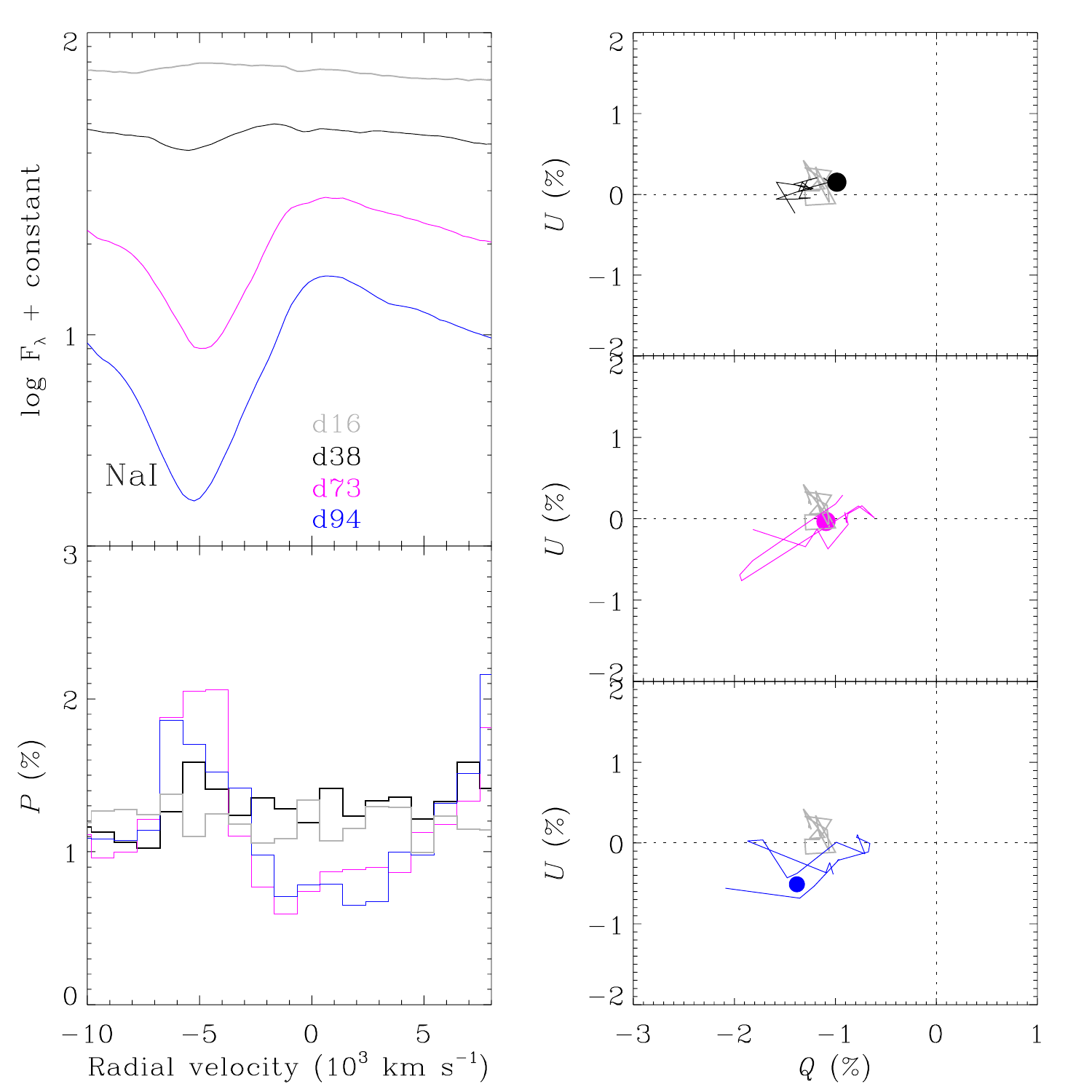}\medskip\\
\medskip
\medskip
\includegraphics[width=3.5in]{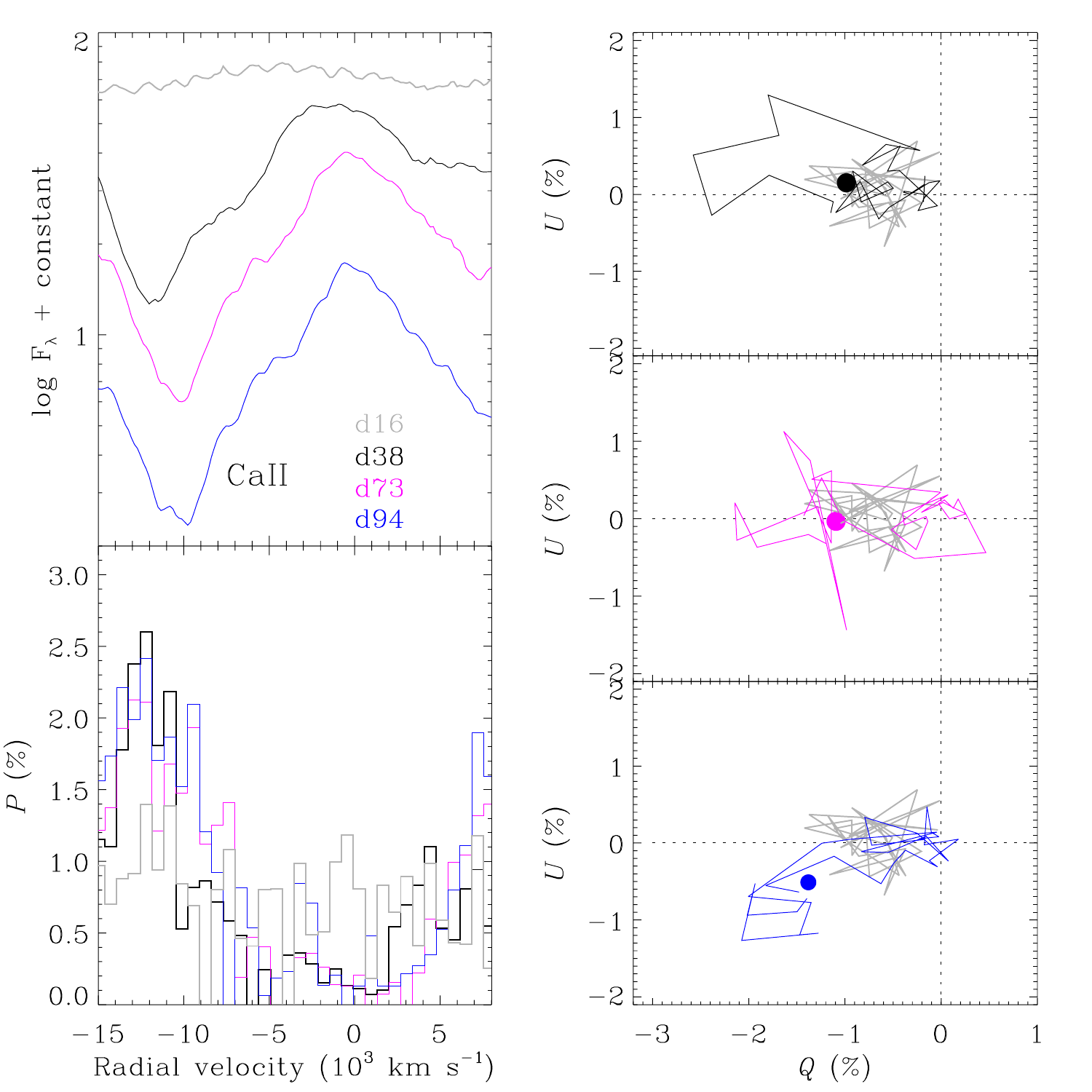}
\includegraphics[width=3.5in]{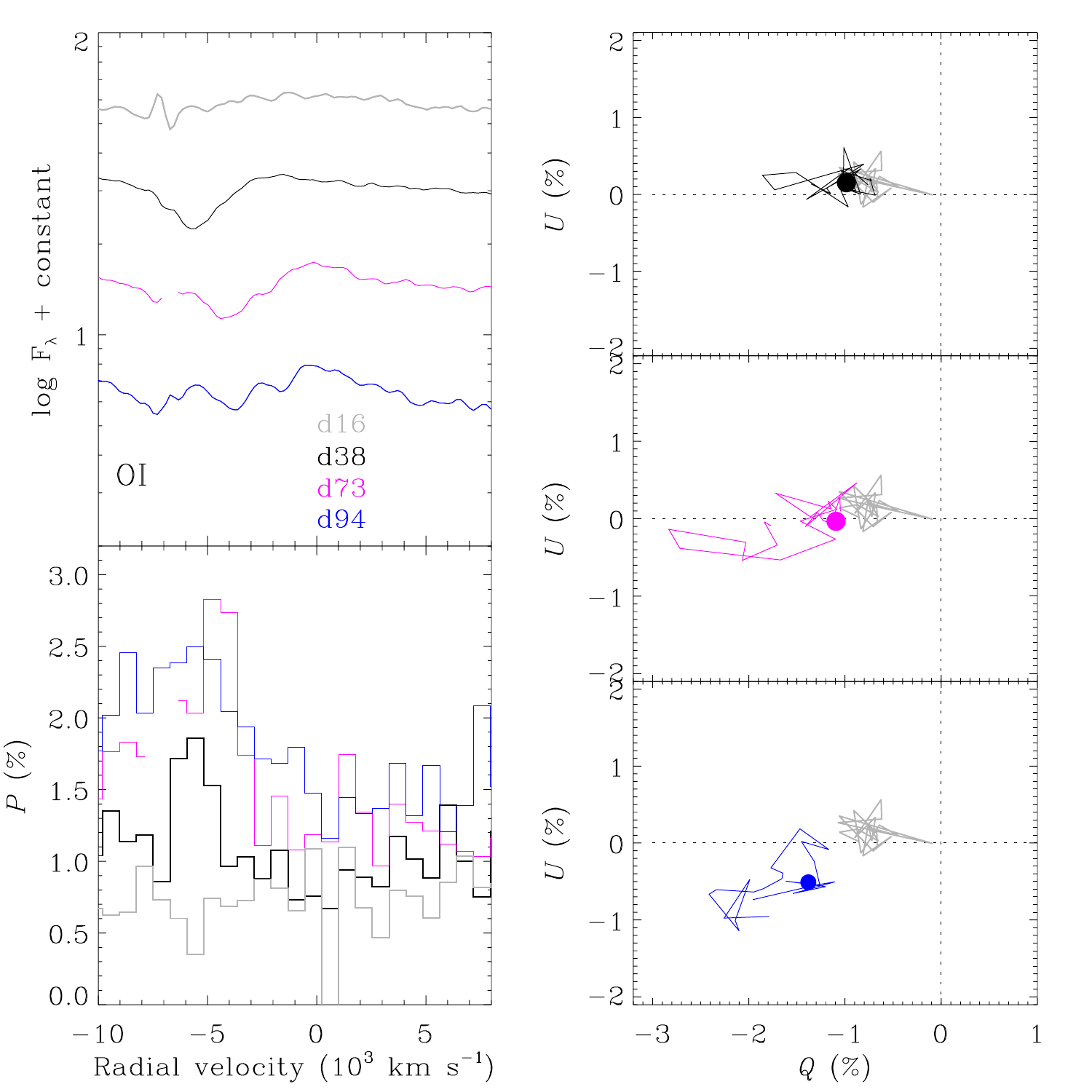}
\caption{Temporal evolution of the H$\alpha$ (upper left), Na\,{\sc i}-D (upper right), Ca\,{\sc ii} IR (lower left), and O\,{\sc i} (lower right) line polarization on days 16 (gray), 38 (black), 74 (magenta), and 94 (blue). For each plot, the upper-left panel shows the flux spectrum, the lower-left panel displays the polarization, and the right three panels show the behavior in the $Q$--$U$ plane. The filled dots in the $Q$--$U$ plane mark the integrated continuum polarization for that date; the day 16 data are shown in all panels for reference. For Na\,{\sc i}, if the feature is instead associated with He\,{\sc i}, then the inferred velocities are $\sim800$\,km\,s$^{-1}$ less.} 
\label{fig:linpol}
\vspace{3mm}
\end{figure*}

By day 19, enhanced polarization across the absorption components of line features starts becoming prominent. The strongest line feature at this stage appears to be associated with H$\beta$, peaking near $\sim2.0$\%. Weaker enhancements are seen for the Ca\,{\sc ii} IR triplet and O\,{\sc i} near 7650\,\AA. There is an absorption line just blueward of the H$\alpha$ component, together appearing as a double-dipped feature. The bluer feature has been identified by Valenti et al. (2014) as Si\,{\sc ii} $\lambda$6355. We observe a slight enhancement in $P$ by $\sim0.3$\% for this feature, relative to the polarization level on either side. Leonard et al. (2013) noted that the appearance of this potential Si\,{\sc ii} feature at such an early phase is unusual, suggesting peculiar ionization conditions not standard for an RSG envelope; this issue will be revisited in \S5. Leonard et al. (2013) also considered the alternative possibility that this feature is a separate high-velocity component of H$\alpha$; however, there is no comparable high-velocity feature associated with H$\beta$. Therefore, we favor the Si\,{\sc ii} identification, which is supported by the spectral modeling of Bose et al. (2015) and Dhungana et al. (2016), and is also the consensus among other studies of SN\,2013ej in the literature (Valenti et al. 2014; Huang et al. 2015; Yuan et al. 2016).

By day 38, the line features in the flux spectrum have strengthened, with substantial deepening of their absorption components. H$\alpha$ absorption begins exhibiting enhanced polarization, with clear depolarization across the emission component. Ca\,{\sc ii} IR polarization dominates the spectrum, peaking near $\sim3.0$\%, with complete depolarization of its emission component. Interestingly, it is at this epoch that the Ca\,{\sc ii} absorption develops multi-component structure. O\,{\sc i} is also enhanced at $\sim2.0$\%. Less prominent line features of Na\,D $\lambda\lambda$5890, 5896, the Sc\,{\sc ii} multiplet near 5700\,\AA, and Fe\,{\sc ii} $\lambda\lambda$5535, 5169 also exhibit enhanced polarization. The overall characteristics remain similar on day 44, but with the Fe\,{\sc ii} polarization climbing above 2.0\%.

Some interesting changes occurred by day 70. Both the line and continuum polarization have further increased by a few tenths of a percent, with strong, broad modulations in polarization becoming prominent, particularly in the region between 5800\,{\AA} and 6500\,{\AA}. The global position angle also appears to have rotated slightly, as illustrated by Figure\,\ref{fig:pt_lc}. Just 3 days later at day 73, highly polarized line features grow narrower in width. He\,{\sc i} features become discernible across much of the optical spectrum. H$\alpha$ line polarization is particularly striking at this epoch, with peak polarization exceeding $\sim3.0$\% and a very narrow morphology; this strongly peaked feature appears to be coincident with the emergence of a asymmetric notch on the blue side of the absorption component, which will be shown in more detail below in Figure\,\ref{fig:linpol}. It is remarkable that the large modulations in polarization, particularly in the wavelength range 6000--6500\,{\AA}, exhibit such constancy in position angle. 

By day 94, H$\alpha$ and H$\beta$ polarization have diminished in strength, while line features associated with Ba\,{\sc ii} $\lambda$6142 and the Sc\,{\sc ii} multiplet near 5700\,{\AA} have increased to almost $\sim3.0\%$.  The continuum polarization at this epoch has increased to its peak level for our coverage at $\sim1.5$\%; however, we note that our chosen continuum sample region also appears to contain a small but non-negligible contribution from line polarization. At this epoch the H$\alpha$ profile develops a blue bump, which is  also shown in more detail in Figure\,\ref{fig:linpol}. 

On day 101 the continuum polarization has dropped back down to $\sim1.2$\%, while only minor changes in line polarization are seen. Just 6 days later, on day 107, the Ba\,{\sc ii} and Sc\,{\sc ii} features around 6100--6200\,{\AA} are no longer discernible, but the continuum polarization has only marginally decreased to $\sim1.0$\%. The line-blanketed region in the blue near $\sim 5000$\,{\AA} appears to drop to null values of polarization.  H$\alpha$ and the Sc\,{\sc ii} multiplet near 5700\,{\AA} remain prominent, however. 

Figure\,\ref{fig:pt_lc} (left panel) illustrates the temporal evolution of the integrated broadband polarization and the position angle. During the first 20 days there is a $\sim0.2$\% fluctuation in $P$ at nearly constant $\theta$.  Subsequently, and through day 44, the continuum and $R$-band $P$ remain within a range of 0.8\% to 1.1\%, while the $V$ band exhibits higher values in the range 1.2\% and 1.3\%. By day 70, the values of all bands approach convergence near $P\approx1.4\%$, but quickly drop again all together by several tenths of a percent only a few days later; $\theta$ also makes a substantial shift at this point and continues to drift thereafter. Specifically, on day 94, when the light curve is near the edge of the plateau, the continuum $P$ and $\theta$ reach their largest excursions from the average, shifting by much more than the $V$ or $R$ bands. This implies that there might be a changing source continuum $P$ that is separate from the line $P$. However, $\theta$ never deviates by more than $\sim13^{\circ}$ at any point. As the light-curve plateau drops down to the radioactive decay tail, the continuum $P$ also drops by several tenths of a percent, while $\theta$ shifts $\sim14^{\circ}$ back toward the values measured on day 73 and earlier. 

Figure\,\ref{fig:pt_lc} (right panel) shows the same data after subtracting the slightly larger value of ISP derived by Kumar et al. (2016). As described in \S2.3, we adopted their $R$-band ISP value and scaled to the $V$-band and continuum regions according to the functional form of Serkowski et al. (1975) and Whittet et al. (1992). The $R$-band imaging polarimetry measurements from Kumar et al. (2016) are also included in the Figure\,\ref{fig:pt_lc} for comparison. We note that their first two epochs of polarimetry near day 100 are significantly below the persistently high polarization trend exhibited by our data, although the error bars associated with their second epoch overlap with our inferred trend. Their third epoch $P$ measurement is consistent with the trend exhibited by our data, but their associated $\theta$ value appears offset by $\sim30^{\circ}$. Their following two measurements past day 110, during the nebular phase, exhibit a significant rise in polarization beyond the temporal range of our coverage. Given the relatively large uncertainties in their measurements, only their first epoch might appear to be inconsistent with the trend inferred from our data. However, the apparent evolution that one might glean from their data alone --- a weakly polarized recombination phase exhibiting a continuous increase in polarization after the plateau, similar to some previous studies of SNe~II-P (e.g., Leonard et al. 2006; Chornock et al. 2010) --- would be inconsistent with the persistently strong polarization we have measured for SN\,2013ej throughout the entirety of the recombination phase.
 
\subsubsection{Behavior in the $Q$--$U$ Plane}
The behavior of the most prominent polarized line features in the $Q$--$U$ plane, and their strength as a function of radial velocity, are illustrated in Figure\,\ref{fig:linpol} for days 16, 38, 73, and 94. On day 16, H$\alpha$ forms a cluster of points exhibiting slight elongation along the $-Q$ axis, with maximum deviation of $P\approx0.8$\% occurring at a radial velocity near $\sim 10,000$\,km\,s$^{-1}$. On day 38 the feature is stronger ($\sim1.8$\%) and clearly elongated along the symmetry axis defined by the continuum region. Peak polarization is near $\sim 10,000$\,km\,s$^{-1}$. By day 73 the feature becomes a tight linear feature and is still aligned with the continuum symmetry axis. Strengthened polarization near 2.3\% is seen out to velocities of $\sim10,000$\,km\,s$^{-1}$, but a prominent ($\sim3.1$\%) narrow peak has developed at $-$8000\,km\,s$^{-1}$. This is the epoch where the high-velocity notch first mentioned in the previous section has appeared in the absorption component. By day 94, the strong feature has weakened only slightly and remains near the same position angle, but has broadened to lower radial velocities of $-$5000\,km\,s$^{-1}$, where the position angle shifts slightly, following the small drift of the continuum axis. At this epoch a bump has emerged on the blue side of the H$\alpha$ profile. 

For Ca\,{\sc ii} IR, the behavior in the $Q$--$U$ plane is quite different from that of H$\alpha$. On day 38, strongly enhanced polarization forms a relatively wide loop in the Stokes plane; maximum polarization still occurs near the position angle of the continuum symmetry axis, but polarization at lower velocities exhibits a $\sim25^{\circ}$ departure from axial symmetry on the sky ($\sim50^{\circ}$ excursion in the Stokes plane). The maximum polarization of Ca\,{\sc ii} also occurs at significantly higher radial velocities than H$\alpha$, peaking near $\sim-12,000$\,km\,s$^{-1}$. By day 73, maximum polarization remains near the same axisymmetric angle, but the position angle at lower velocities exhibits more erratic behavior in the $Q$--$U$ plane. But by day 94, the Ca\,{\sc ii} feature forms a more elongated structure in the plane that extends into the same position angle as that of H$\alpha$ and the other lines in Figure\,\ref{fig:linpol}, but at significantly higher velocities. 

For the Na feature, maximum polarization occurs on day 73, as it does for H$\alpha$. The line also forms a linear structure in the $Q$--$U$ plane at this epoch. The position angle is comparable to that of H$\alpha$, but is slightly shifted by about $10^{\circ}$ on the sky. The velocities associated with the enhanced polarization are significantly lower than in the case of H$\alpha$, peaking near $-5000$\,km\,s$^{-1}$ (assuming association with Na\,{\sc i}), and they likely originate at deeper layers in the homologous outflow.

O\,{\sc i} appears to follow an evolutionary trend similar to that of other lines, and at velocities comparable to those of Na\,{\sc i}. However, by day 94 the development of  multiple absorption components in the flux spectrum suggests that contamination by other lines might be an issue for this later epoch.

\subsection{Spectropolarimetric Modeling}

\subsubsection{Dust-Scattering Model}
SN\,2013ej exhibited X-ray evidence for interaction with tenuous CSM (Chakraborti et al. 2016), the density of which is consistent with the mass loss of a typical RSG wind. Moreover, as we will show in \S3, CSM interaction persists at late times, implying the presence of a highly extended tenuous CSM envelope. RSG winds have long been known to contain dust (Zuckerman 1980), and high-resolution imaging and polarimetric observations of Betelgeuse and Antares have shown that such dust can be clumpy (Kervella et al. 2011, 2016; Ohnaka 2014); this material could provide an asymmetric scattering medium and, if viewed at a suitable orientation, a major source of polarization. Depending upon the composition and size of dust grains and their distance from the SN, circumstellar dust can potentially survive if the UV pulse of SN shock breakout fades faster than the sublimation timescale at a given radius. The polarizing effect of dust scattering should therefore be considered in the case of SN\,2013ej.

Prior to Wang \& Wheeler (1996), strong, broad modulations in polarization across line features were taken as evidence against circumstellar scattering; those authors showed that interpretation to be incorrect, following their re-analysis of the polarization data from SN\,1987A. Indeed, the observed polarization depends on both the intrinsic SN light curve and the scattering process in a complicated way (see Wang \& Wheeler 1996, their Equation 3). The net signal is a mix of direct light from the SN, with highly polarized scattered light from an earlier spectroscopic phase, echoed by distant CSM dust.  In short, the net polarization develops wavelength dependence because the luminosity evolves differently for different wavelengths; this is especially true for P-Cygni features. Therefore, polarization modulations across lines are not exclusive to electron scattering in the SN envelope. 

\begin{figure}
\includegraphics[width=3.3in]{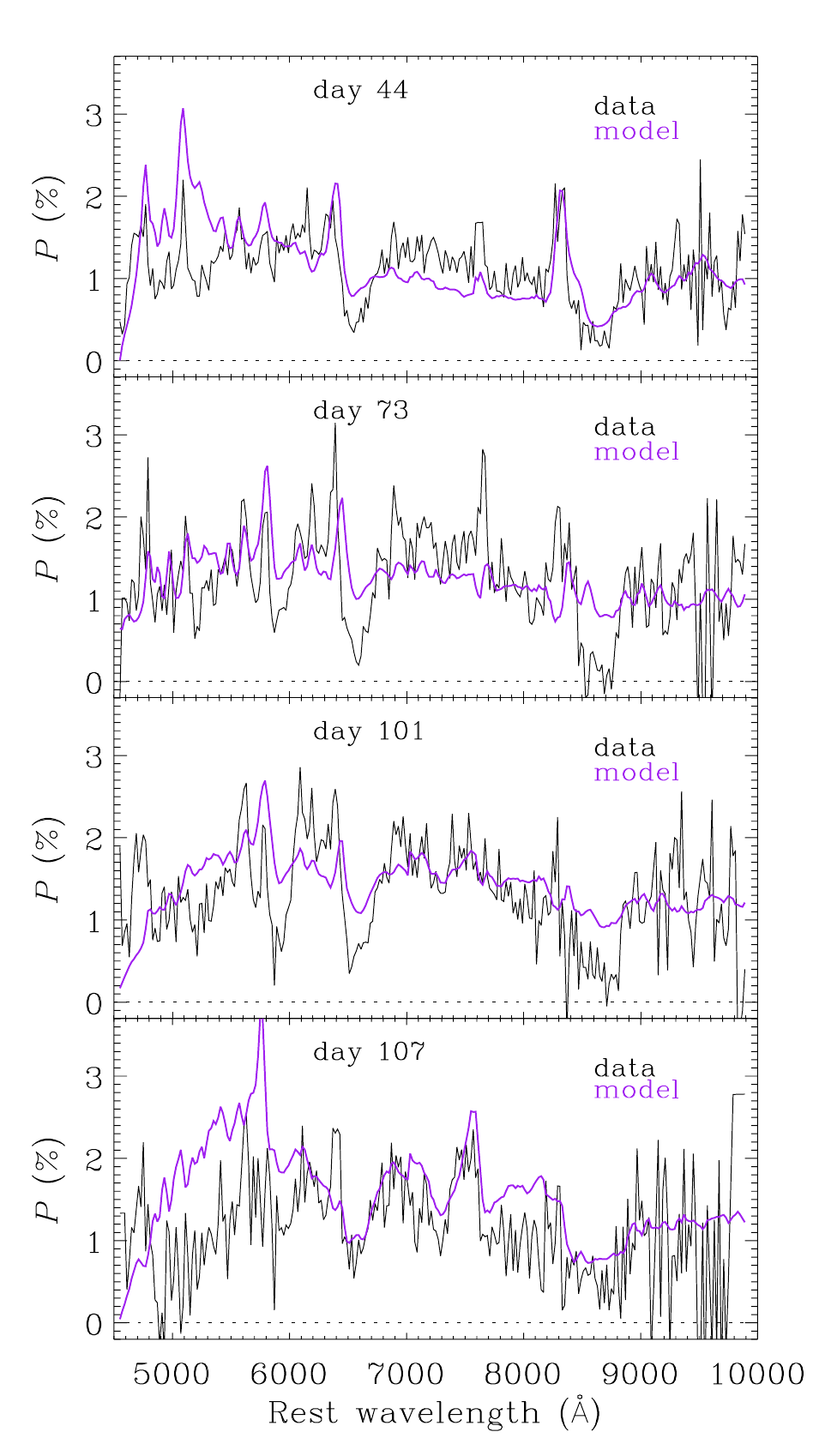}
\caption{The polarization spectrum $P$ from dust-scattering spectropolarimetry models (purple) compared to the data (black) at the observed epochs. We refer the reader to Figure\,\ref{fig:p_theta} for line identifications.}
\label{fig:dustpol}
\vspace{3mm}
\end{figure}

We therefore constructed a model for which the polarized signal is the result of optical scattering by dust particles in the CSM, building upon the methodology of Wang \& Wheeler (1996). The detected photons consist of those propagating directly from the SN mixed with those scattered toward the observer by the circumstellar dust. This mode of polarization does not require intrinsic asphericity of the SN. The basic elements of the model, which will be presented in more detail in a forthcoming paper (M. Hu et al., in preparation), are summarized as follows. The available flux spectra were used as the input source of light, and the luminosity evolution was calibrated using the observed light curve (for this we used all of the spectra and photometry originally presented by Huang et al. 2015). The asymmetric component of CSM was modeled as a single stationary sphere of dust with radius $1.3\times10^{16}$\,cm, displaced from the SN by $7.8\times10^{16}$\,cm in the plane of the sky ($\sim30$ day light-travel time from the SN), resulting in a $90^{\circ}$ scattering angle for the observer; this orientation is a reasonable approximation, since photons that scatter at right angles will dominate the polarized signal. The models were constructed by trial-and-error to provide the best overall visual matches to the multi-epoch data. In the end, the separation of $7.8\times10^{16}$\,cm for the dust sphere was used because it provided the best match to the observed spectropolarimetric evolution.  Model calculations were performed and are presented for $t=44$, 73, 101, and 107 days. The optical depth to dust scattering is allowed to vary from epoch to epoch, and the best-matching models were obtained for optical depths that decreased continually from values of 0.1 to 0.05 between 44 and 107 days, respectively. The model does not distinguish between graphite and silicate grains, although we note that graphite grains are much more likely to survive the SN at radii within $<10^{17}$\,cm for grain sizes $<0.1\,\mu$m (see Wang \& Wheeler 1996, their Figure\,1).

The synthetic polarized spectra from our dust-scattering models are compared with the data in Figure\,\ref{fig:dustpol}.  Considering its geometric simplicity, the model reproduces many of the observed features in the polarized spectrum reasonably well. For day 44, the model generates substantial continuum polarization and strongly polarized line features. In particular, Ca\,{\sc ii} IR line polarization is closely matched in terms of strength and radial velocity. The model H$\alpha$ line exhibits similar morphology and blueshift relative to the data, but does not fully reproduce the depth of the polarization trough associated with the emission component. The Fe\,{\sc ii} feature near wavelength 5000\,{\AA} and H$\beta$ at the blue end of the spectrum are generated by the model, although at significantly higher strengths than we observed. The weak O\,{\sc i} $\lambda$7774 is also fairly well reproduced on this date in terms of strength and radial velocity.

On days 73 and 101, the observed $\sim$1\% continuum level is well matched. However, the model fails to reproduce the polarized line profile of Ca\,{\sc ii} IR, showing a relatively narrow trough in polarization where an enhancement is observed. The model does generate line features that correspond with most of those that are present in the data; however, the features are not matched well in strength. A particularly noteworthy aspect of the modeling at these epochs is the relative blueshifts of polarized line features. While the model Na\,{\sc i}\,D, Sc\,{\sc ii}, Fe\,{\sc ii}, and Ba\,{\sc ii} features match the data in radial velocity, the H$\alpha$, O\,{\sc i}, and Ca\,{\sc ii} features in the data exhibit substantially higher radial velocities. We speculate that this discrepancy might provide a means to distinguish between multiple components of line polarization (i.e., those intrinsic to the SN and those produced by scattering in CSM). 

By day 107, the model match improves for some of the broad modulations across the polarized spectrum, while worsening for some of the narrow line features. Specifically,  the depth of the broad H$\alpha$, O\,{\sc i}, and Ca\,{\sc ii} appears well matched, while the model exhibits a strong Na\,{\sc i}\,D polarization feature that is no longer present in the data at this epoch. The model also predicts the vanishing of the narrow H$\alpha$ feature altogether, yet this feature remains strong in the data on day 107 and thus appears to be better matched by the day 101 model. The blue end of the model spectrum exhibits significantly higher polarization than the data, although the downturning trend in the spectrum toward shorter wavelengths is reproduced.  

Our basic scattering model has substantial limitations in reproducing some of the detailed features and evolution. Perhaps most importantly, the single dust-sphere configuration results in no scattered photons being observed until after the associated light-travel delay time of $\sim30$ days, yet strong polarization is indeed observed from the earliest epoch on day 11. A more realistic model might involve multiple clumps in a more continuous distribution of circumstellar dust. Furthermore, 
pure dust scattering will always produce some polarized flux at the wavelengths of the strong emission peaks in the flux spectrum. In this regard, the model best matches the data from day 107, after the recombination phase has ended, and, to a lesser extent day 44. However, on day 73, in particular, we see that H$\alpha$ and Ca\,{\sc ii} emission exhibit nearly complete depolarization, which suggests that electron scattering is likely to be the dominant mode of scattering on these epochs.

\subsubsection{Electron-Scattering Model}
We experimented with the construction of electron-scattering models to match the flux and polarization spectra from day 73, which is the epoch where the line polarization is most pronounced and is not satisfactorily reproduced by the dust-scattering models presented above. This epoch also exhibits nearly complete depolarization of the H$\alpha$ and Ca\,{\sc ii} emission peaks, which is another indication that electron scattering denominates the polarization. We used the program {\tt synpol}, which was utilized for the spectropolarimetric analysis of the Type\,Ia SN\,2001el in the spectral vicinity of the Ca\,{\sc ii} IR feature (see Kasen et al. 2003 for details); {\tt synpol} is an extension of the program {\tt  synow} (e.g., Branch et al. 2007), which is used for modeling flux spectra of SNe. For polarization modeling, {\tt synpol} allows for an ellipsoidal photosphere, geometrically represented as an inner unpolarized boundary surface surrounded by a pure electron-scattering envelope having a power-law electron-density profile of the form $\rho\propto r^{-2}$. The program {\tt synpol} does not satisfactorily reproduce the strong, broad emission features in the flux spectra of SNe, because, like {\tt synow}, the code assumes local thermodynamic equilibrium (LTE) and thus does not account for collisional excitation, which is important in the line-forming regions of SNe. Consequently, {\tt synpol} is not able to fully reproduce the broad dips in polarization associated with the emission components of the flux spectrum, which result from the fact that the line-forming region is optically thin to electron scattering and adds a source of mostly unpolarized flux to the net signal.

\begin{figure}
\includegraphics[width=3.3in]{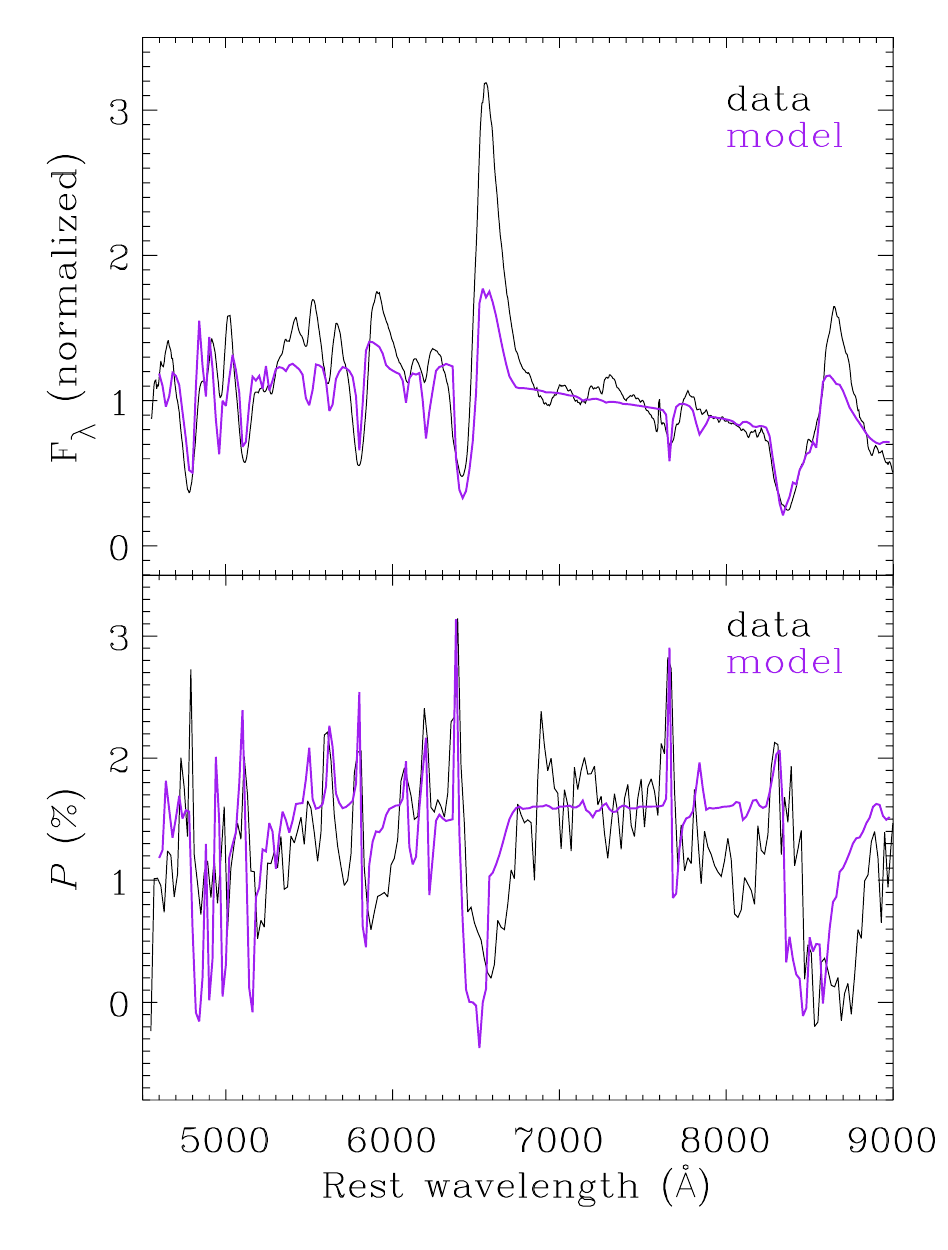}
\caption{Our best-matching {\tt synpol} model for the normalized flux (upper panel) and polarization spectrum (lower panel) of SN\,2013ej on day 73. We refer the reader to Figure\,\ref{fig:p_theta} for line identifications.}
\label{fig:synpol}
\vspace{3mm}
\end{figure}

Uneven absorption of polarized photospheric light by ions in the overlying line-forming region can produce enhanced polarization features at the wavelengths of absorption components in the flux spectrum. From the observer's perspective, the fastest radial velocity region on the approaching side of the homologous outflow will preferentially obscure the central portion of the underlying photosphere. Since photospheric light emerging from that central region is mostly forward scattered, it does not contribute substantially to the net polarization of the unresolved source. The blueshifted absorption thus removes the weakly polarized light from the net signal, while strongly polarized light from the photospheric limb (where radially diffusing photons will be scattered at angles near $\sim90^{\circ}$) emerges. This effect thus increases the fractional polarization of the blueshifted side of the absorption profile, blueward of the flux minimum. This mode of line polarization will produce line polarization exhibiting a high degree of axisymmetry, and will produce linear features in the $Q$--$U$ plane, such as those we see for H$\alpha$ on day 73 (see Figure\,\ref{fig:linpol}); however, this mode requires that the underlying photosphere is aspherical, in order to prevent net cancellation of all electric vectors emerging from the photospheric limb photons. The alternative case of patchy line absorption by clumps of ions over a spherical photosphere can also produce polarized lines, but in this case the features will be non-axisymmetric, and will produce circular loops in the $Q$--$U$ plane --- for example, more consistent with the relatively high-velocity Ca\,{\sc ii} feature on day 38 in Figure\,\ref{fig:linpol}.

The {\tt synpol} program allows for various geometric configurations for the line-forming region, including ellipsoids, spherical clumps, and toroids. A homologous flow is assumed, whereby the radial extents of the photosphere and line-forming regions are defined by velocity coordinates, set by the user. For clumps of line absorption randomly distributed across an aspherical photosphere, one would expect to observe substantial modulations in $\theta$ across the wavelength range of the spectral absorption features, since both axisymmetric and non-axisymmetric components of the photosphere will be obscured over the associated range of radial velocities. For similar reasons, a toroidal distribution of line absorption will also exhibit substantial shifts in $\theta$ with respect to radial velocity, particularly if the toroid is viewed nearly edge-on. For SN\,2013ej, the observed high degree of axisymmetry between the polarized absorption features and the continuum appears be more consistent with an ellipsoidal geometry that matches the ellipsoidal geometry of the underlying photosphere. 

\begin{figure*}
\includegraphics[width=3.8in]{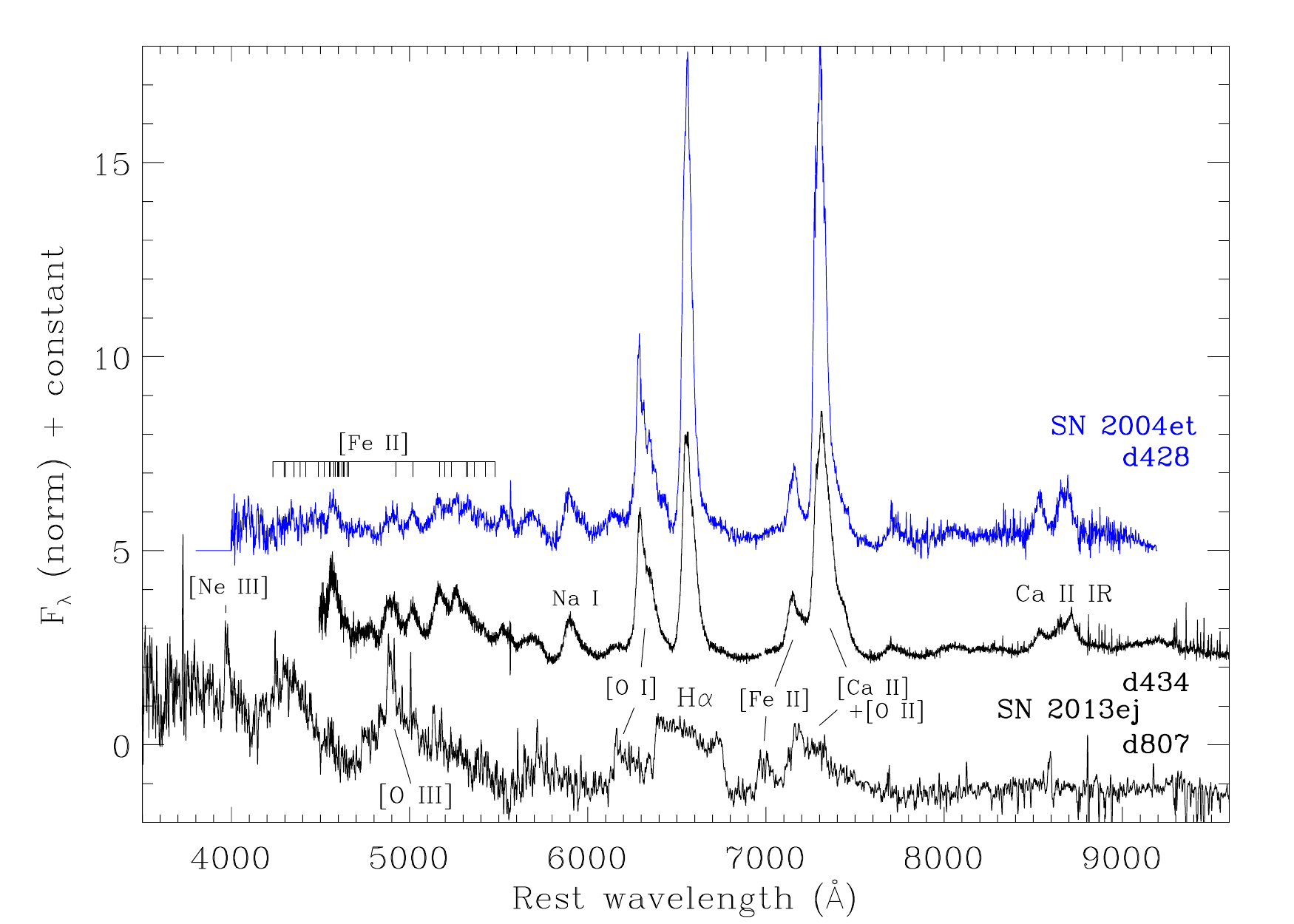}
\includegraphics[width=2.7in]{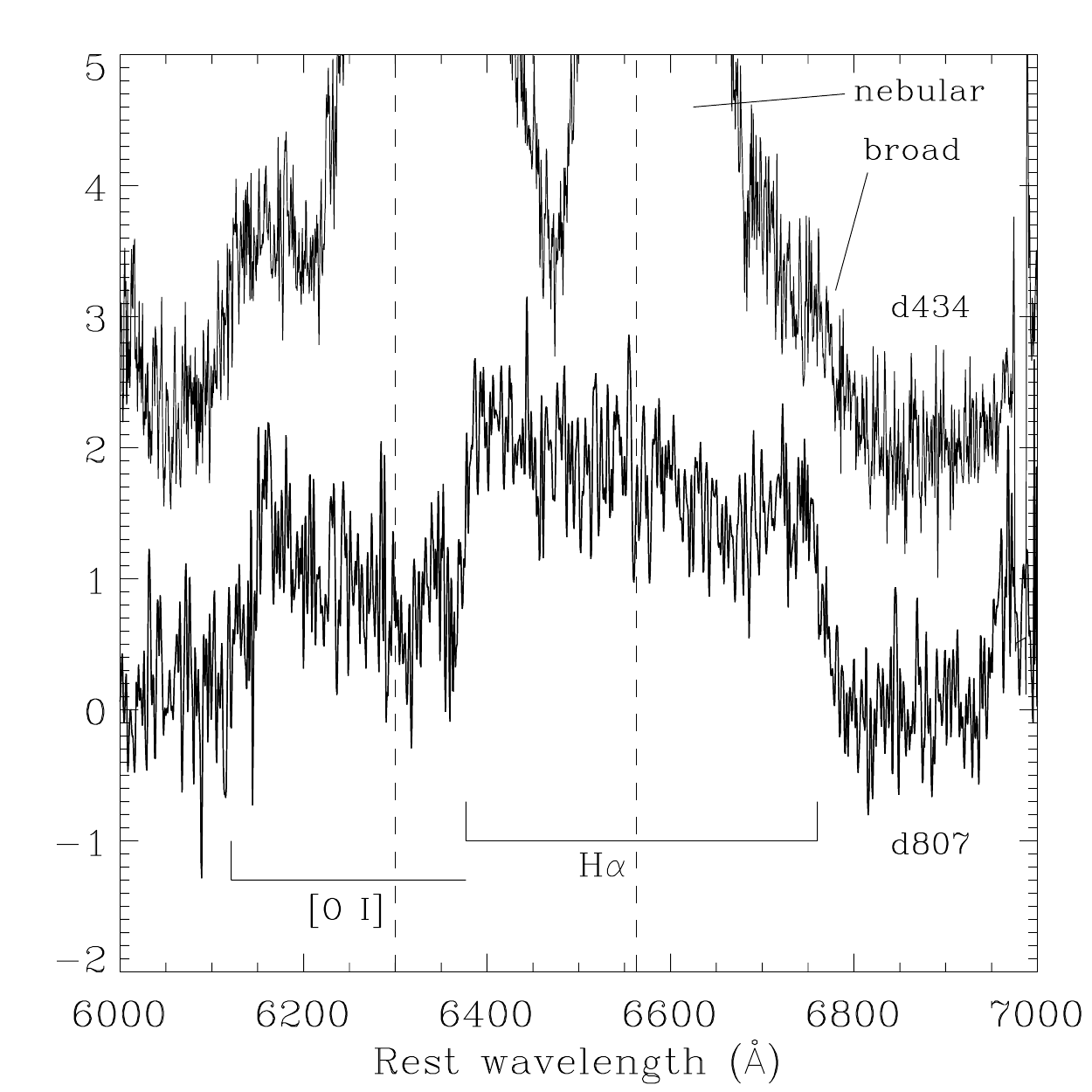}
\caption{Left panel: late-time Keck/LRIS spectra of SN\,2013ej (black) from days 434 and 807. A comparison spectrum of the Type II-P SN\,2004et (blue; Sahu et al. 2006) on day 428 is also shown in the left panel for comparison, which was obtained from the WISeREP database (Yaron \& Gal-Yam 2012). Right panel: an expanded view of the region near H$\alpha$ and O\,{\sc i}. The boxy profile on day 807 appears to have been present on day 434 as well, underneath the strong nebular features.}
\label{fig:late_spec}
\vspace{3mm}
\end{figure*}

We therefore constructed our synthetic polarization spectra using an ellipsoidal photosphere and line-forming region. The following atomic transitions were included in the model: H\,{\sc i}, O\,{\sc i}, Na,{\sc i}, Mg\,{\sc i}, Si\,{\sc ii}, Ca\,{\sc ii}, Fe\,{\sc ii}, Ti\,{\sc i}, Ti\,{\sc ii}, and Ba\,{\sc ii}.  The radii and optical-depth profiles of the line-forming envelopes were also chosen by trial and error until a satisfactory match to the polarized features was obtained. The power-law exponent, $n$, of the radial density profile, $r^{-n}$, of the line-forming region was set to have $n=8$ for all transitions. However, reproducing the strongly peaked polarized line profile of H$\alpha$ required $n=6$. The best matches to the flux and polarization spectrum were obtained with blackbody and excitation temperatures of 5500\,K and a photospheric radius of 4100\,km\,s$^{-1}$ (velocity coordinates). We note that matching the strongly peaked H$\alpha$ line required setting the line optical depth, $\tau$, to a value of 150, whereas most of the other line matches were obtained with optical depths in the range 5--10, with the exception of Ca\,{\sc ii}, which required $\tau=220$ to get the best match. 

Our {\tt synpol} model for the flux and polarized spectrum is presented in Figure\,\ref{fig:synpol}. The best overall match was found using an ellipsoidal photosphere 
with an axis ratio of 1.4, viewed at 90$^{\circ}$ inclination angle (edge-on). Slightly higher values of axis ratio also produced consistent matches if the inclination was slightly decreased. However, our best matches occurred for axis ratios in the range 1.3--1.5 at inclination angles within $\pm10^{\circ}$ of edge-on. The strongly peaked polarized line features were matched reasonably well using line-forming regions that are attached to the photosphere (radii beginning at velocity coordinate 4100\,km\,s$^{-1}$) and with the same ellipsoidal geometry (spherical line-forming regions worked similarly well), with the exception of Ca\,{\sc ii} IR, for which we obtained our best match using a detached line-forming region with inner radius at 5400\,km\,s$^{-1}$; this necessary detachment might be related to the non-axisymmetric geometry of this feature exhibited in Figure\,\ref{fig:linpol}. 

There are large discrepancies between the data and model emission components of the flux spectrum. This is to be expected, since, as stated earlier, our calculations assume LTE and do not account for collisional ionization, which contributes to the emission-line flux. The depolarizing effect of collisional ionization also results in deeper polarization troughs than our model produces, which if taken into account would probably improve the match.

The ellipsoidal model produces purely axisymmetric line polarization, roughly consistent with the high degree of axisymmetry we measure (i.e., the $\lesssim10^{\circ}$ drift in $\theta$) across most of the polarized line. Realistically, small deviations from a perfectly aligned photosphere and line-forming region could likely account for the minor spread in $\theta$ observed across the spectrum. However, the erratic structure of  Ca\,{\sc ii} in the $Q$--$U$ plane (see Figure\,\ref{fig:linpol}) indicates non-axisymmetric asphericity for the absorption associated with that ion, which our preliminary model does not explain.

Overall, the {\tt synpol} model provides a reasonable match to the day 73 data, and thus demonstrates that electron scattering in an ellipsoidal geometry viewed nearly edge-on could explain the continuum polarization, and the morphologies of most of the narrowly peaked polarized line features and their associated radial velocities. However, the ability for both electron-scattering and dust-scattering models to generate many of the general spectropolarimetric features of SN\,2013ej indicates that both processes could potentially contribute to the observed characteristics. If the position angles of dust- and electron-scattering components are similar, the constructive interference of their associated Stokes vectors could lead to an overestimate of the axis ratio of the electron-scattering photosphere.

\begin{figure}
\includegraphics[width=3.2in]{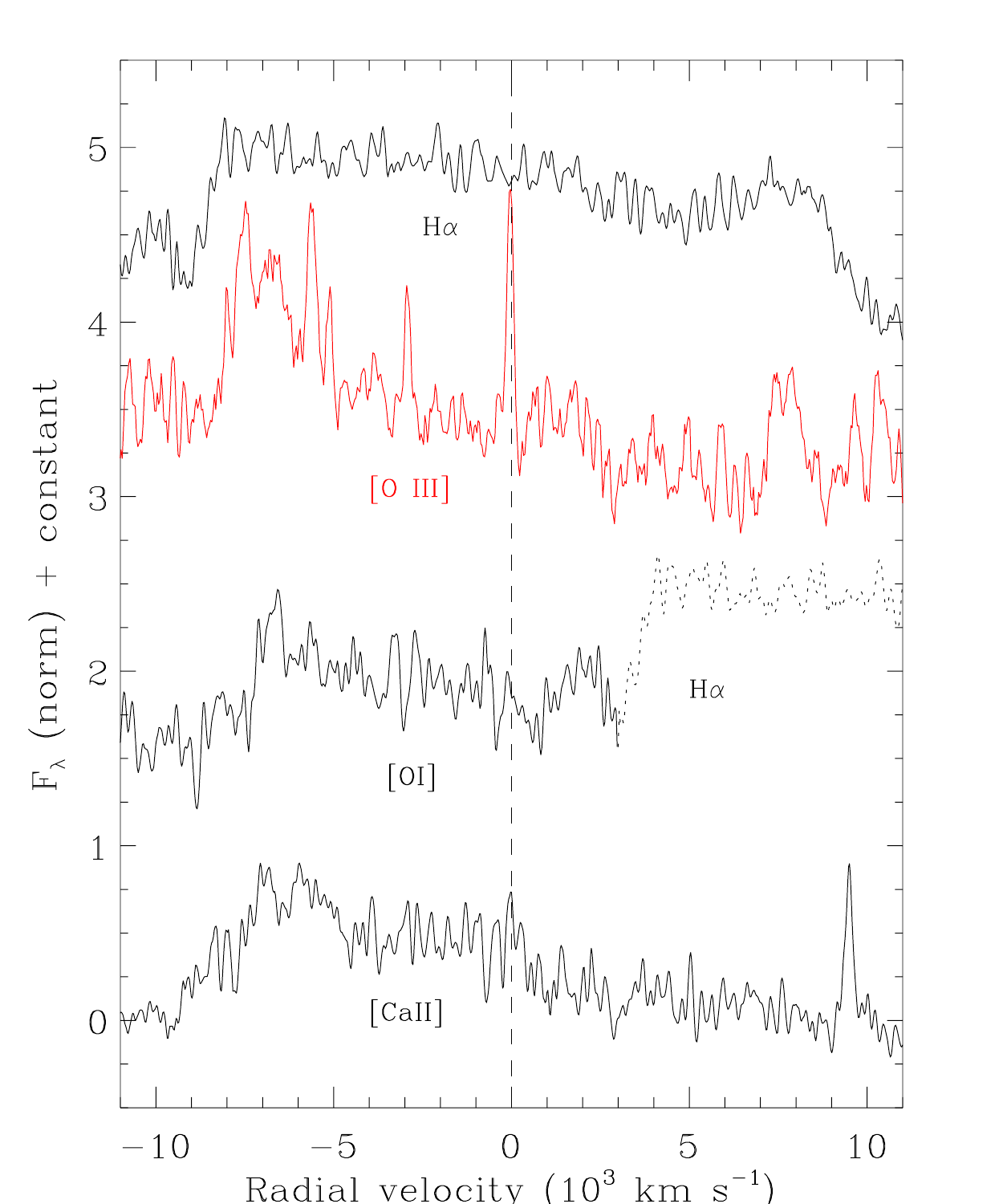}
\caption{Radial-velocity profiles of selected emission lines in the Keck/LRIS spectrum of SN\,2013ej from day 807. }
\label{fig:rvline}
\vspace{3mm}
\end{figure}

\begin{figure}
\includegraphics[width=3.2in]{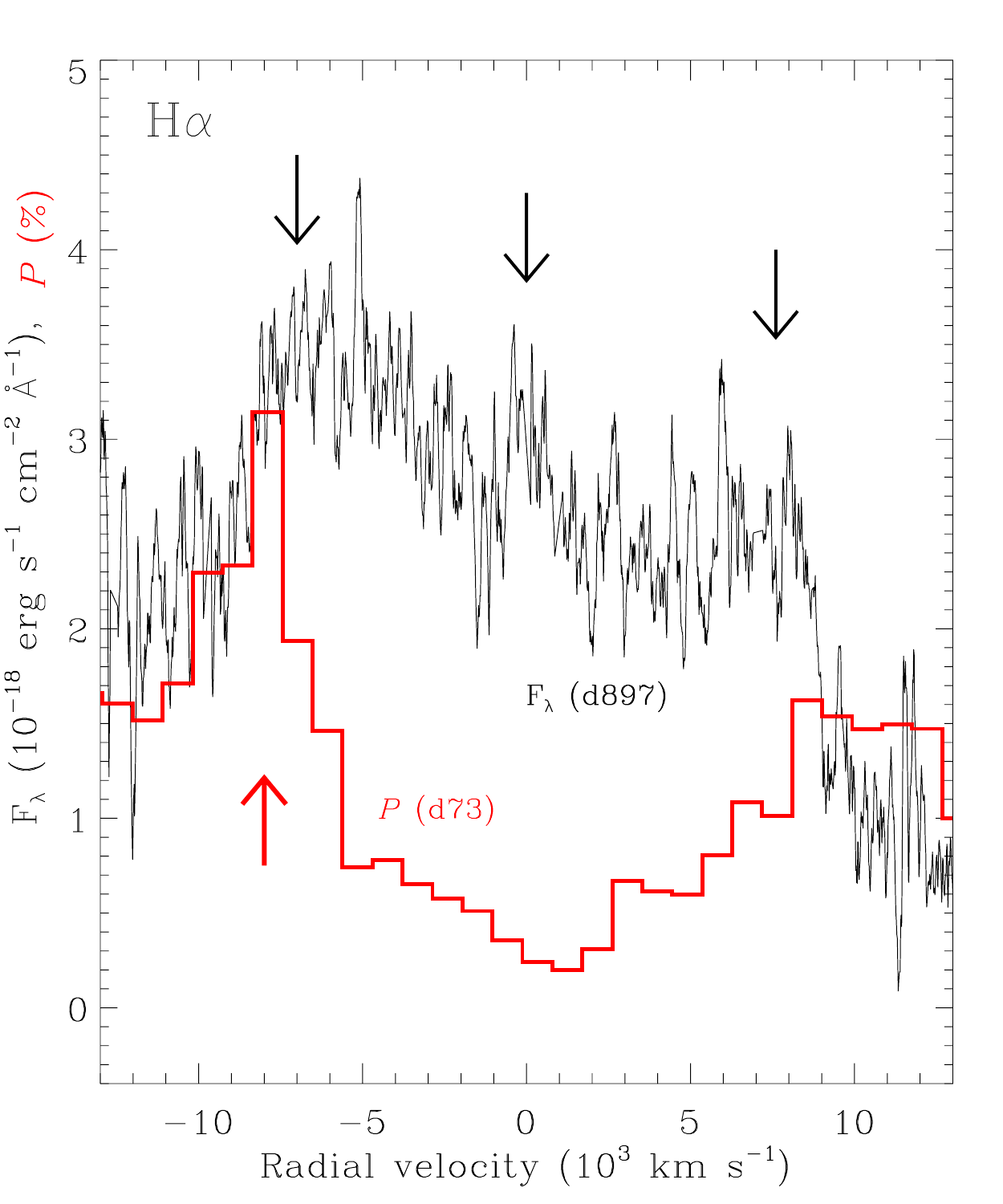}
\caption{Keck/DEIMOS spectrum of SN\,2013ej from day 897 (black). The peak of line polarization from day 73 (red) is included to highlight the similar velocities of the polarization feature and the edge of the late-time flux profile. Arrows mark regions of interest, including possible discrete individual emission components within the multi-peaked flux profile.}
\label{fig:ha_velpol}
\vspace{3mm}
\end{figure}

\section{Deep Late-Time Spectroscopy}
Late-time optical spectroscopy of SN\,2013ej was obtained on 2014\,Oct.\,2 (day 434) and 2016\,Jan.\,8 (day 897) at Keck Observatory using the Deep Imaging Multi-Object Spectrograph (DEIMOS; Faber et al. 2003), and on 2015\,Oct.\,10  (day 807) using the Low-Resolution Imaging Spectrometer (LRIS; Oke et al. 1995). The day 434 DEIMOS data were obtained with the 600ZD grating and 1$\farcs$0 slit for a FWHM resolution of $\sim3.5$\,{\AA}; two exposures were obtained for a total integration time of 1200\,s. The day 807 LRIS observations were performed using the 600/400 grism for the blue channel and the 400/8500 grating for the red channel; three exposures were obtained for a total integration time of 3600\,s . The day 897 DEIMOS data were taken with the 1200\,l\,mm$^{-1}$ grating and 0$\farcs$8 slit for a FWHM resolution of 0.9--1.3\,{\AA}; five exposures were obtained for a total integration time of 7500\,s. The DEIMOS and LRIS observations at all epochs were performed at an airmass of 1.0--1.1 and with the slit aligned with the parallactic angle (Filippenko 1982). All data were flat-fielded and wavelength calibrated using standard techniques. Flux calibration was applied using observations of standard stars at similar airmass, and telluric lines were removed.

The spectra from days 434 and 807 are presented in Figure\,\ref{fig:late_spec}. On day 434, SN\,2013ej exhibits a late nebular-phase emission spectrum, dominated by strong lines of H$\alpha$, [O\,{\sc i}], [O\,{\sc ii}], [Ca\,{\sc ii}], [Fe\,{\sc ii}], Na\,{\sc i}, and Ca\,{\sc ii} IR. The spectrum closely resembles that of SN\,2004et on day 428 of its evolution (Sahu et al. 2006), and is characteristic of a nebular-phase spectrum from a core-collapse SN.

By day 807 the spectrum of the SN has changed dramatically. H$\alpha$ line emission exhibits a very broad and boxy profile, much broader than the nebular spectrum from a year prior, with a bumpy multi-peaked structure. There is also a broad asymmetric emission line of [O\,{\sc I}] $\lambda\lambda$6300, 6364 which overlaps with the blue side of H$\alpha$, in addition to a blend of [O\,{\sc ii}] $\lambda$7325  and [Ca\,{\sc ii}] $\lambda\lambda$7291, 7324, and emission from [Fe\,{\sc ii}] $\lambda$7155. Closer inspection of the earlier day 434 spectrum, facilitated by the right-hand panel of Figure\,\ref{fig:late_spec}, shows that the same broad, weak components of H$\alpha$ and [O\,{\sc i}] from day 807 were already present on day 434, underlying the stronger, narrower nebular line profiles.  There is weak emission likely to be associated with Na\,D or He\,{\sc i} $\lambda$5876. Blueward of that feature the spectrum forms a pseudocontinuum resulting from a blend of multiple emission lines, including [Fe\,{\sc ii}] transitions and prominent [O\,{\sc iii}] $\lambda$5007.

\begin{figure}
\includegraphics[width=3.45in]{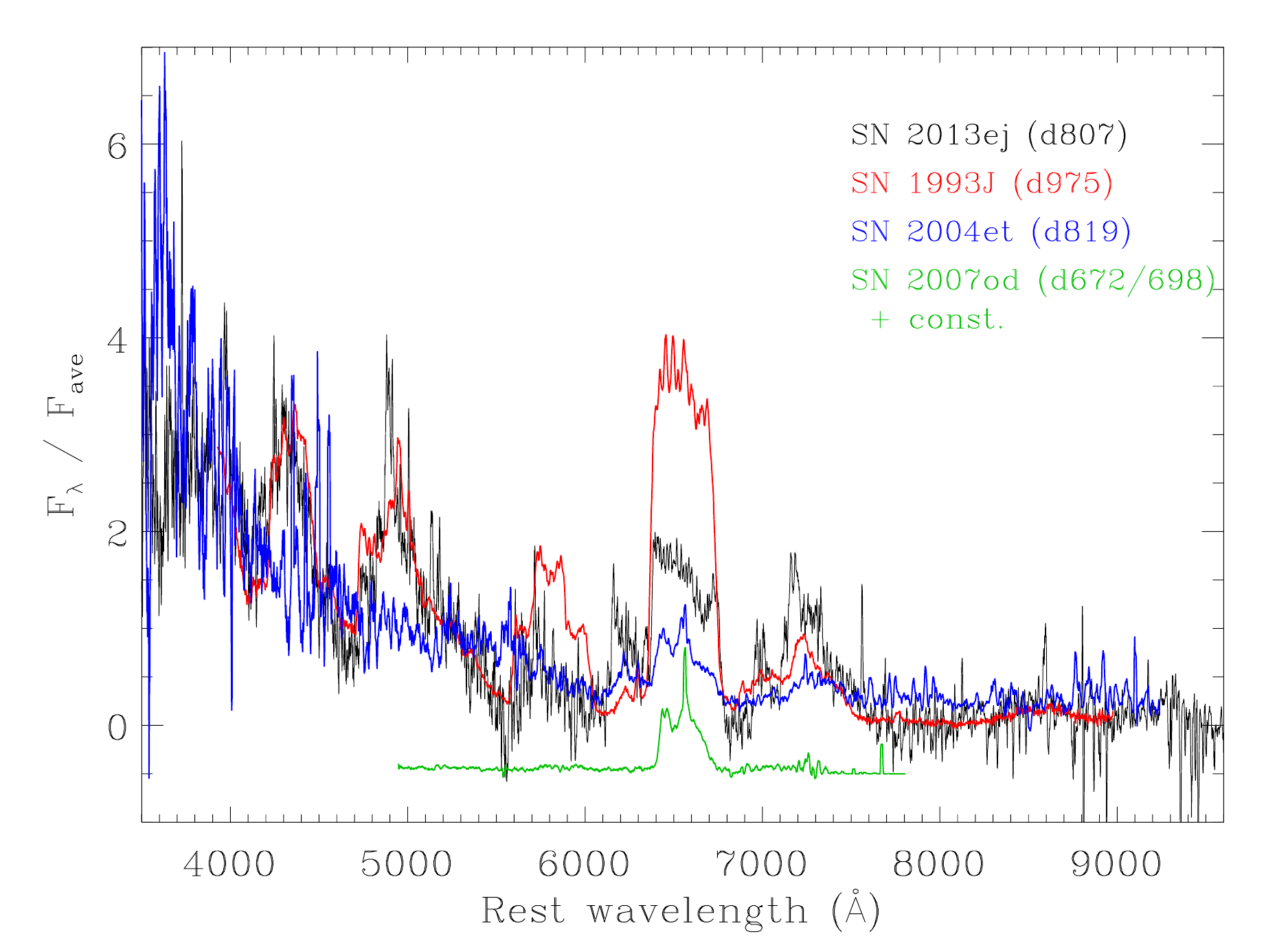}
\caption{Day 807 spectrum of SN\,2013ej (black) compared to late-time spectra of SN\,1993J (red; Matheson et al. 2000), SN\,2004et (blue; Faran et al. 2014), and SN\,2007od (Andrews et al. 2010). The spectra have been normalized by their average values (i.e., divided by a constant). The spectrum of SN\,2007od was further scaled and offset by a constant for clarity.}
\label{fig:latespec_compare}
\vspace{3mm}
\end{figure}

All of the broad emission lines exhibit strongly blueshifted peaks.  Figure\,\ref{fig:rvline} shows an expanded view of H$\alpha$, [O\,{\sc iii}], [O\,{\sc i}], and [Ca\,{\sc ii}] plotted as a function of radial velocity (day 807). The H$\alpha$ profile exhibits steep edges extending out to $\sim9000$\,km s$^{-1}$ on the blue side and $\sim10,000$\,km s$^{-1}$ on the red side. There appears to be preferential suppression of flux on the red side of the profile, with the exception of enhanced emission at the far-red edge. The same overall line morphology is apparent for [O\,{\sc iii}] $\lambda$5007, but with stronger red-flux suppression.  For this line there also appears to be a strong emission component on the red side near $+8000$\,km\,s$^{-1}$, although this might be emission from a different line. A very narrow component of nebular [O\,{\sc iii}] emission is seen at rest velocity, and is probably from the host H\,{\sc ii} region. [O\,{\sc i}] and [Ca\,{\sc ii}] also exhibit line morphologies with blueshifted peaks, although the velocity widths appear to be slightly smaller than those of H$\alpha$ and [O\,{\sc iii}].

The higher-resolution DEIMOS spectrum of H$\alpha$ from day 897 is shown in Figure\,\ref{fig:ha_velpol}, along with the polarized line profile on day 73, when the line polarization has reached maximum. The flux-profile morphology is roughly consistent with the LRIS spectrum from day 807, but the higher-resolution data exhibits more pronounced detail. In particular, there is possible indication of a triple-peaked structure, including a central intermediate-width (FWHM $\approx 1000$ km s$^{-1}$) component near rest velocity, bounded by two broader components. Interestingly, the blue edge of H$\alpha$ extends to the same high value of radial velocity as the peak of the polarized line profile from day 73, which suggests that the polarized line feature is an intrinsic feature of SN originating in the outer ejecta; we discuss this further in \S5. 

\begin{figure*}
\centering
\includegraphics[width=5.3in]{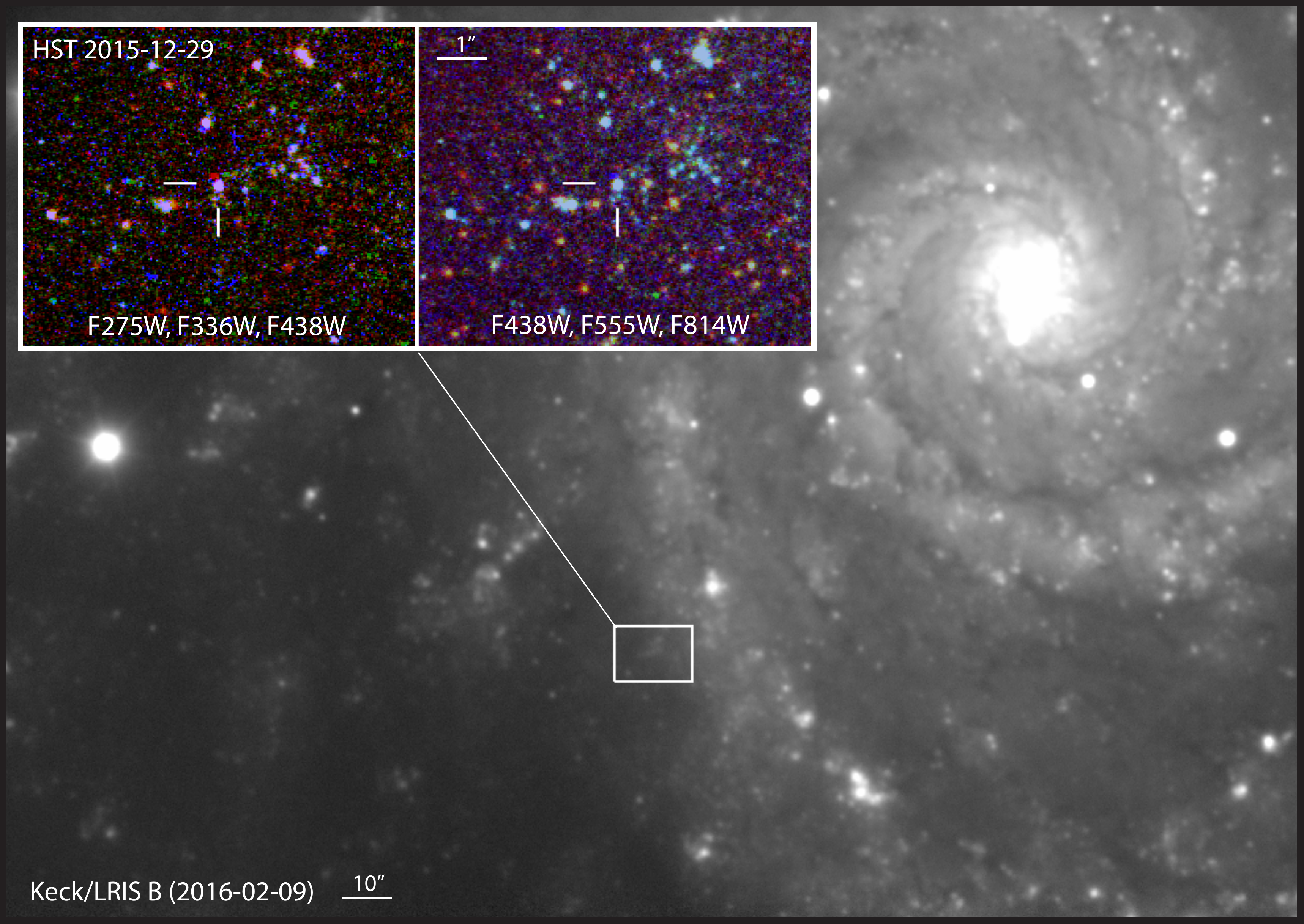}
\caption{Keck/LRIS $B$-band image (black and white background) of SN\,2013ej and M74 on 2015 Feb. 9. The inset shows {\it HST} data from 2015 Dec. 29 used to extract the photometry presented in Table\,2; the color images are a blue-green-red composite constructed from the respective F275W, F336W, F438W (left inset) and F438W, F555W, F814W images (right inset). North is up and east toward the left in all images.}
\label{fig:montage}
\vspace{3mm}
\end{figure*}

The overall appearance of the day 807 spectrum resembles that of the Type IIb SN\,1993J at day 975, shown in Figure\,\ref{fig:latespec_compare}. H$\alpha$ and Na\,{\sc i}/He\,{\sc i} emission, with respect to the other spectral features, are much stronger in SN\,1993J, but the widths of the lines are roughly the same and both exhibit red suppression. Their resemblance is particularly striking in the blue pseudocontinuum, including the broad [O\,{\sc iii}]/H$\beta$ and blended [Fe\,{\sc ii}] features. SN\,2013ej also shares similarities with the Type II-P SN\,2004et (day 819), although the latter does not exhibit the strong emission features in its blue continuum and the H$\alpha$ profile is more asymmetric and less boxy in shape. The relative strength of the blueshifted component of emission from [O\,{\sc i}] $\lambda6300$ and [O\,{\sc i}]$+$[Ca\,{\sc ii}] near 7200\,{\AA} is most pronounced in SN\,2013ej.

The broad, boxy line profiles observed on days 807 and 897 are the result of CSM interaction, generated as the SN ejecta become energized by passage of the reverse shock. The shocked ejecta experience collisional excitation while pre-shock ejecta can be photoionized by X-rays and UV emission from the hot plasma near the reverse shock (Chevalier \& Fransson 1994). Both processes potentially contribute to the broad-line emission, although post-shock collisional excitation is probably the dominant contributor to H$\alpha$. 

CSM interaction through day 897 probes mass loss from the progenitor going back at least $\sim2500$\,yr before core collapse, assuming shock and CSM-wind velocities of 9700\,km\,s$^{-1}$ and 10\,km\,s$^{-1}$, respectively (Chakraborti et al. 2016). Meanwhile, the earlier presence of broad emission on day 434, seen underneath the stronger nebular line profiles, indicates that CSM interaction was occurring at this phase as well. This is consistent with long-term X-ray observations, which showed that SN\,2013ej was indeed persistently interacting with the RSG wind for at least 145 days (Chakraborti et al. 2016). Such highly extended, tenuous, and asymmetric CSM is reminiscent of the environments surrounding the RSGs Antares and Betelgeuse (Kervella et al. 2011, 2016; Ohnaka et al. 2014), and also motivates our construction of dusty-CSM scattering models to investigation the spectropolarimetry (see \S2).

\section{Late-Time Ultraviolet--Infrared Photometry}
\subsection{HST/WFC3}
\begin{table}\begin{center}\begin{minipage}[bp]{3.3in} \setlength{\tabcolsep}{2.8pt}
      \caption{Late-time {\it HST} Photometry of SN\,2013ej.\footnote{MJD 57387 (day 888) adn MJD 57666 (day 1169).}}
\centering
  \begin{tabular}{@{}cccc@{}} 
  \hline
Band     & mag\footnote{The {\it HST} flight-system magnitudes were transformed to $BVI$ following Sirianni et al. (2005). Uncertainties are statistical.}  & flux ($\mu$Jy) & Epoch (day)   \\                            
\hline
\hline
F275W  &  $23.00\pm0.35$  &   $0.58\pm0.19$   &  888   \\  
F336W &  $23.26\pm0.10$  &    $0.61\pm0.06$  & 888 \\ 
F438W  &  $23.70\pm0.03$ &    $1.39\pm0.04$   & 888 \\ 
F555W &  $23.52\pm0.02$  &    $1.45\pm0.03$ & 888 \\
             &  $23.38\pm0.02$  &   $1.65\pm0.03$ &  1169 \\
F814W & $23.03\pm0.03$  &    $1.52\pm0.04$ & 888 \\
             & $22.75\pm0.03$  &    $1.97\pm0.04$ & 1169  \\

\hline
\end{tabular} \end{minipage} \end{center}
\label{hstphot}
\end{table}

\begin{figure}
\includegraphics[width=3.35in]{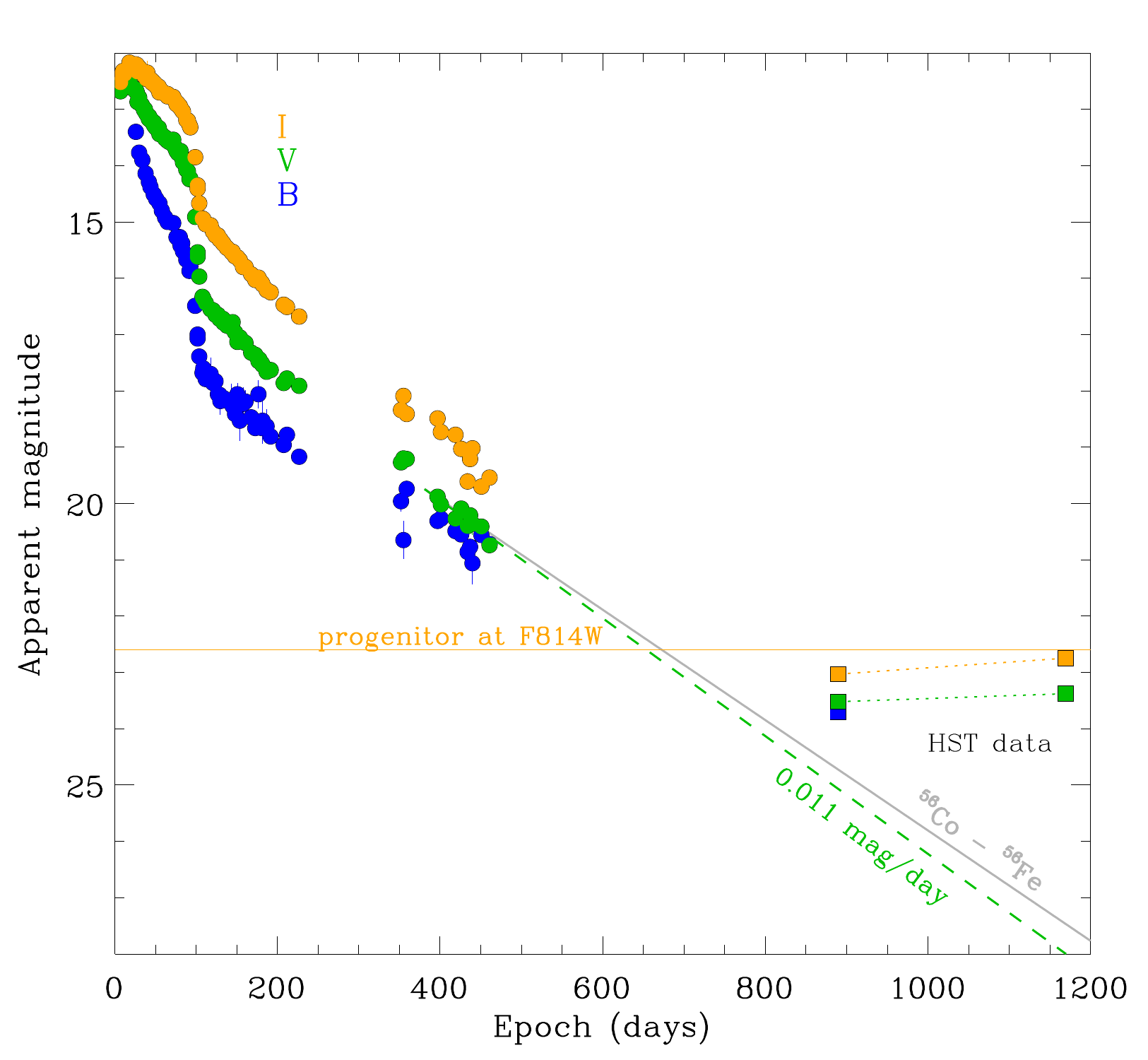}
\caption{{\it HST} $BVI$ photometry from days 888 and 1169 (squares) shown with the earlier light-curve data from Dhungana et al. (2016), whose late-time decline fit of 0.011 mag day${-1}$ is illustrated by the green dashed line. The radioactive $^{56}$Co decay rate is represented by the solid gray line. The {\it HST} F814W ($I$-band) progenitor level is illustrated by the solid orange horizontal line.}
\label{fig:opt_lc}
\end{figure}

\begin{figure*}
\centering
\includegraphics[width=7in]{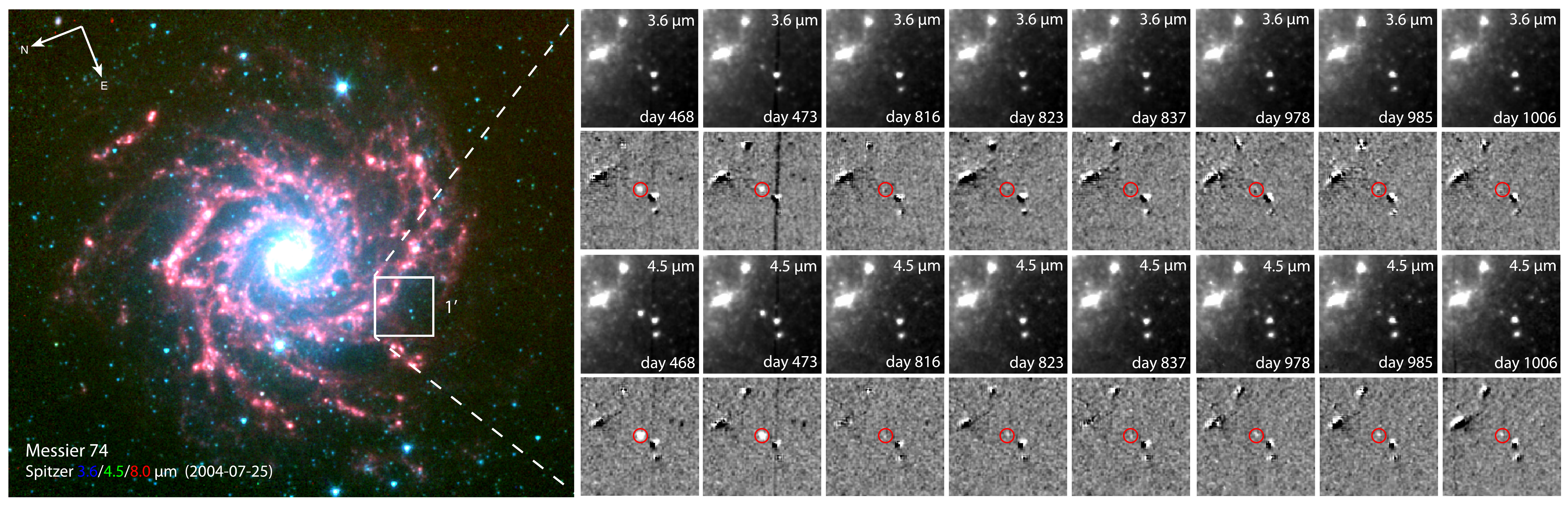}
\caption{RGB composite of the \textit{Spitzer}/IRAC template images of M74 from 2004 (left), and the template-subtracted images of the region around SN\,2013ej (tiled frames, right).}
\label{fig:subs}
\vspace{3mm}
\end{figure*}

High-resolution imaging observations of SN\,2013ej were performed with the {\it HST} and Wide-Field Camera 3 (WFC3) on 2015\,Dec.\,29 (day 888) under {\it HST} program GO-14116 (PI S. Van Dyk) and on 2016\,Oct.\,4 (day 1169) under {\it HST} Snapshot program GO-14668 (PI A. Filippenko).  Exposures were obtained in the F275W ($NUV$), F336W ($U$), F438W ($B$), F555W ($V$), and F814W ($I$) filters on day 888, and only the latter two filters on day 1169. The point source is detected in all available bands from both of these late epochs, shown in the inset of Figure\,\ref{fig:montage}. The aim of the program was to detect a light echo around SN\,2013ej. %No resolved echo is apparent in the images in any of the bands, although the presence of some flux from a compact echo cannot be ruled out. 

Photometry of the source was extracted from the images using Dolphot. We tried two different approaches to estimate the background, including the use of an annulus region to measure the sky (FitSky=1) and, alternatively, measuring the sky within the point-spread function (PSF) aperture (FitSky=3; best to use when the field is very crowded). Our annulus-based background subtraction produced the most consistent results for all bands, although the results from each setting are within the respective uncertainties. The photometry is listed in Table\,2.

Figure\,\ref{fig:opt_lc} shows the long-term light curve of SN\,2013ej from the beginning of the explosion out to day 1169. Photometry of the cool supergiant progenitor of SN\,2013ej, presented by Fraser et al. (2014), was contaminated by a neighboring blue source with $m$(F435W) = 25.05\,mag and $m$(F555W) = 24.74\,mag; the photocenter through those filters was offset from that of the SN. However, the photocenter of the F814W source in pre-explosion imaging, for which Fraser et al. (2014) measured m(F814W) = $22.66\pm0.03$\,mag, accurately matched that of the SN and has therefore been attributed to the progenitor star. We performed Dolphot runs on both of the available pre-explosion F814W epochs from 2003\,Nov.\,20 and 2005\,Jun.\,16, obtaining $22.60\pm0.01$ and $22.57\pm0.02$, respectively (both values are averages from the results of Dolphot runs using FitSky=1 and FitSky=3 background estimation parameters), slightly brighter than the results from Fraser et al. (2014).  Our photometry from days 888 and 1169 shows that the F814W flux of the SN remnant dropped below the faintest recorded state of the progenitor by $\Delta m=0.46\pm0.04$\,mag, confirming the RSG stars's association with the SN. 

Our day 888 and 1169 {\it HST} photometry shows that the SN remnant is substantially brighter than what is expected from both the earlier 0.011\,mag\,day$^{-1}$ optical decline rate measured by Dhungana et al. (2016) and the expected rate of decay of $^{56}$Co to $^{56}$Fe. Moreover, the source has increased slightly in brightness between days 888 and 1169, by $\sim0.14$~mag  and $\sim0.28$~mag in the $V$ and $I$ bands, respectively; the corresponding color change is a slight reddening by $\Delta(V-I)=0.15$~mag. We attribute the slow rate of decline between the last ground-based coverage and the HST observations on day 888, and recent slight rise in flux between days 888 and 1169, to line emission from CSM interaction, which was revealed by our deep late-time spectroscopy presented in \S3. The color change could be due to emission-line evolution or dust extinction (see \S4.3). However, we note the possibility that the photometry in the bluer bands on day 888 might be slightly contaminated by the neighboring blue source identified by Fraser et al. (2014) in the pre-explosion data. The latest available data show that the SN remnant is still $\gtrsim1$ mag brighter than the blue neighboring source, so any contamination from the neighboring source is probably not substantial in our most recent measurements.  Future epochs of {\it HST} photometry will help disentangle the SN emission from neighboring sources more accurately, once the emission from the SN and CSM interaction fades.

\subsection{Keck/LRIS}
We obtained a $B$-band image of SN\,2013ej on 2016 Feb. 9 with Keck/LRIS. Under 1{\arcsec} seeing conditions, we were unable to resolve the SN emission from
neighboring sources. We thus rely on our higher-resolution {\it HST} images for our late-time optical photometry, but we show the SN position in the host galaxy with the $B$-band LRIS image in Figure\,\ref{fig:montage}.

\subsection{Warm \textit{Spitzer}}
SN\,2013ej was observed between days 238 and 1006 during the \textit{Spitzer Space Telescope} Warm Mission utilizing channels 1 (3.6\,$\mu$m) and 2 (4.5\,$\mu$m) of the Infrared Array Camera (IRAC; Fazio et al. 2004). We acquired fully coadded and calibrated data from the {\it Spitzer} Heritage Archive\footnote{http://sha.ipac.caltech.edu/applications/Spitzer/SHA/.}, including data from Program IDs 11063, 11053, and 10046 (PIs Kasliwal, Fox, and Sanders, respectively). The first three epochs of photometry of SN 2013ej were reported by Tinyanont et al. (2016). We extracted photometry from all epochs independently, registering each SN image with earlier pre-SN images of the host galaxy, which were used as subtraction templates. The template-subtracted images are shown in Figure\,\ref{fig:subs}. We performed aperture photometry on the template-subtracted (PBCD / Level 2) images using a 6-pixel aperture radius and aperture corrections listed in Table\,4.7 of the \textit{Spitzer} IRAC Instrument Handbook\footnote{http://irsa.ipac.caltech.edu/data/SPITZER/docs/irac/iracinstrumenthandbook/27/}. The background and noise levels in the subtracted images were computed from 100 randomly placed apertures near the SN location.The photometry results are listed in Table\,3. 

\begin{figure}
\includegraphics[width=3.2in]{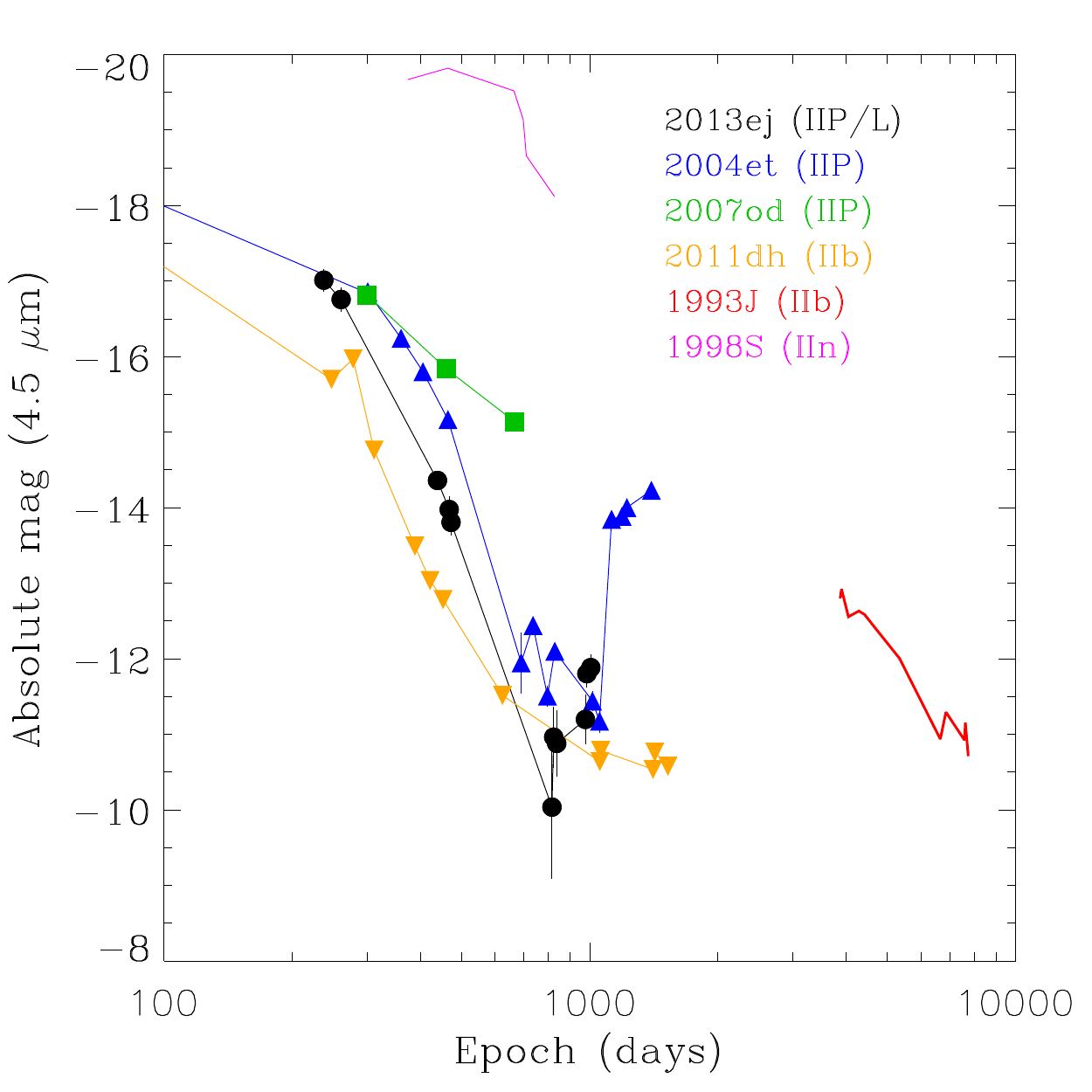}
\caption{Long-term mid-IR light curve of SN\,2013ej (black), and comparison objects SN\,2004et (blue; Kotak et al. 2009), SN\,2007od (green; Andrews et al. 2010), and SN\,1993J (red; Tinyanont et al. 2016). Data for SN\,2011dh (yellow; Helou et al. 2013; Ergon et al. 2015) and SN\,1998S (magenta; Gerardy et al. 2002; Pozzo et al. 2004) are also included for comparison.}
\label{fig:late_midir_lc}
\vspace{3mm}
\end{figure}

The SN is detected at 4.5\,$\mu$m in all epochs up to day 1006 (see light curve in Figure\,\ref{fig:late_midir_lc}). In the 3.6\,$\mu$m data, however, the SN is visible only until day 472,  after which we conservatively list the flux as 3$\sigma$ upper limits in Table\,3.

The mid-IR light curve of SN\,2013ej is consistent with the general behavior of other core-collapse SNe (e.g., see Tinyanont et al. 2016). It most closely resembles the mid-IR light curves of SN\,2004et, which was optically classified as a SN~II-P (Li et al. 2005), and SN\,2007od, which was also classified as a SN~II-P (Andrews et al. 2010), but potentially classifiable as a SN~II-L, based on the criterion of Anderson et al. (2014). Interestingly, both SNe exhibited evidence for late-time CSM interaction, similar to that of SN\,2013ej (see Figure\,\ref{fig:latespec_compare}). Based on their optical to IR SEDs, the mid-IR counterparts of SN\,2004et and SN\,2007od have been shown to be consistent with emission from warm dust (Kotak et al. 2009; Andrews et al. 2010; Inserra et al. 2011).  Late-time \textit{Spitzer} observations of SN\,2004et showed that the mid-IR flux increased dramatically after day 1000, interpreted as the result of dust formation within a dense post-shock shell that developed in response to the CSM interaction (Kotak et al. 2009). There is marginal evidence of a similar late-time increase developing in SN\,2013ej; this is an interesting possibility, given the complementary evidence for CSM interaction at this late stage, and motivates continued monitoring.  

  \begin{table}
      \caption{Spitzer Photometry of SN\,2013ej.}
\centering
  \begin{tabular}{@{}lcccc@{}} 
  \hline
 MJD     & Epoch (day)  & 3.6\,{$\mu$m} ($\mu$Jy)\footnote{Uncertainties are statistical.} & 4.5\,{$\mu$m} ($\mu$Jy)$^\textrm{a}$  & $M_{4.5}$ \\                            
\hline
\hline
 56735.0 & 237.6  & $       765.2 \pm 99.3 $  &$  1420.1 \pm 196.1 $      & $-17.00 $ \\
 56758.3 & 260.9  & $ 	586.7 \pm 69.1 $   &$   1122.1 \pm 168.1 $ & $-16.75$ \\
 56936.6	& 439.2  & $      38.4 \pm 6.7 $     &$    123.5 \pm 13.7 $   & $-14.35$  \\
 56965.3 & 467.9  & $	24.1 \pm 8.9 $     &$     86.6 \pm 13.9 $  & $-13.97$  \\
 56970.1 &  472.7  & $	26.1 \pm 12.6 $    &$     74.1 \pm 12.0 $   &  $-13.80 $  \\
 57313.0 & 815.6  & $	<6.2  $      &$       <5.3 $   &  $-10.03$  \\
 57320.5 & 823.1  & $	<8.9  $      &$       5.4 \pm 2.1 $  &   $-10.95$  \\
57334.2 & 836.8  & $	<8.1 $      &$       5.0 \pm 2.1 $  &  $-10.87$ \\
57474.9 & 977.5  & $	<7.3 $      &$       6.7 \pm 2.0 $   &  $-11.19$  \\
57482.4	& 985.0  & $      <8.6 $      &$     11.7 \pm 2.3 $    & $-11.79$ \\
57503.6 & 1006.2  & $	<9.9 $      &$     12.6 \pm 2.1 $  &  $-11.87$ \\
\hline
\end{tabular}
\label{table:spitzer_phot}
\vspace{3mm}
\end{table}

\begin{figure}
\includegraphics[width=3.2in]{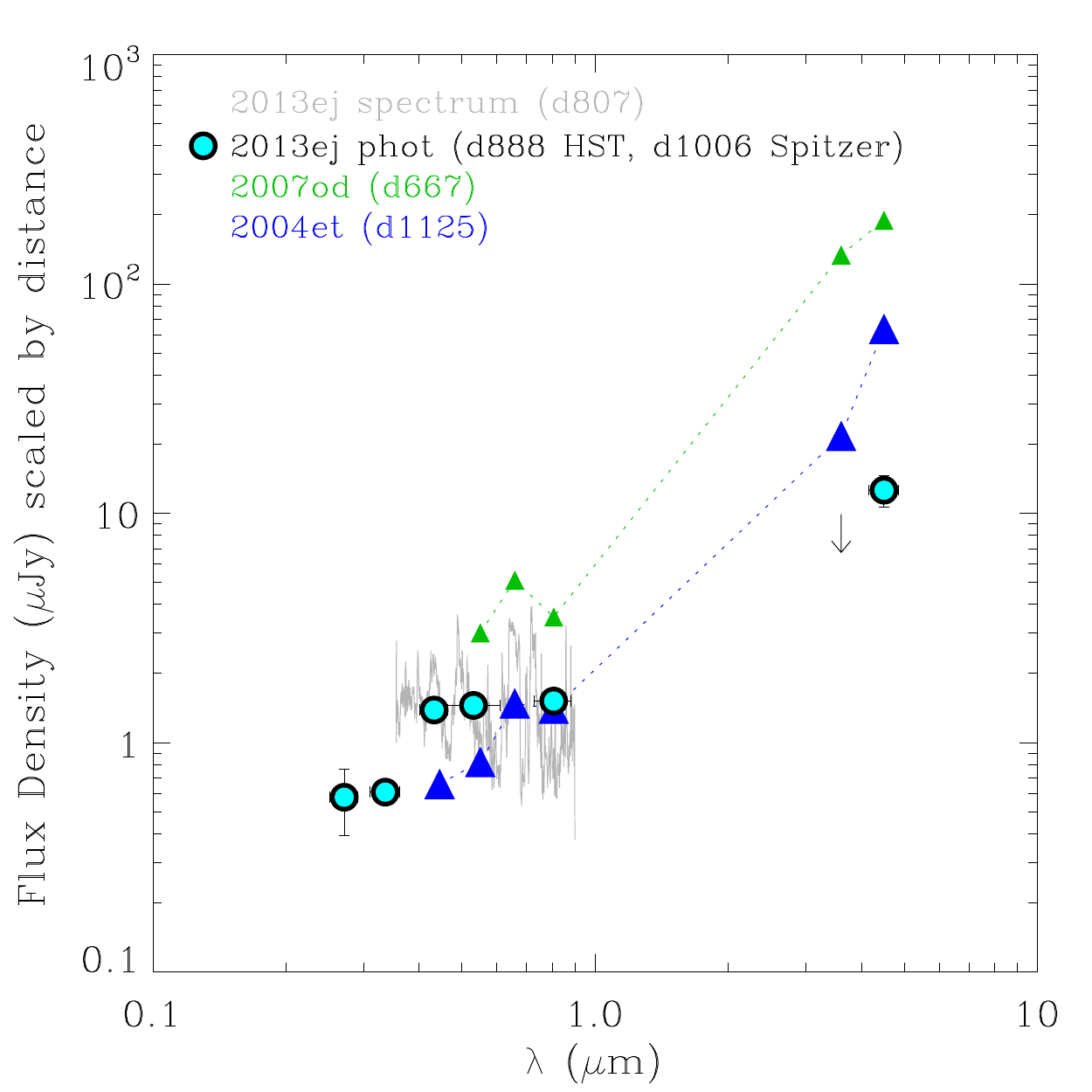}
\caption{Late-time optical through IR SED for the remnant of SN\,2013ej, using the {\it HST} and \textit{Spitzer} data listed in Tables\,2 and 3. The day 807 Keck spectrum is included (gray curve). The SEDs of SN\,2004et (Kotak et al. 2009) and SN\,2007od (Andrews et al. 2010) are included for comparison, and their fluxes have been scaled to the distance of SN\,2013ej.}
\label{fig:late_midir_sed}
\end{figure}

The UV--IR spectral energy distribution (SED) of SN\,2013ej is shown in Figure\,\ref{fig:late_midir_sed}, combining the {\it HST} photometry from day 888 with the \textit{Spitzer} photometry from day 1006, when the late 4.5\,$\mu$m flux has slightly rebrightened.  The broadband optical photometry from day 888 is more or less consistent with the Keck/LRIS spectrum from day 807. The IR excess from the SN, like the mid-IR light curve, appears similar to that of SN\,2004et and SN\,2007od. The source of the excess is likely to be thermal, although it is also possible that the 4.5\,${\mu}$m flux is also influenced by line emission.

\subsubsection{Dust Emission Model}
We analyze the \textit{Spitzer} data using a simple dust model (Fox et al. 2011). We assume that the source of the mid-IR excess is warm dust, and fit the SED as a function of the dust mass ($M_{\rm d}$) and temperature ($T_{\rm d}$). The flux is given by (Hildebrand 1983)
\begin{equation}
\label{eqn:flux2}
F_\nu = \frac{M_{\rm d} B_\nu(T_{\rm d}) \kappa_\nu(a)} {d^2},
\end{equation}
where $a$~is the dust grain radius, $\kappa_\nu(a)$ is the dust absorption coefficient, and $d$ is the distance of the dust from the observer.  We assume $a=0.1\,\mu$m and derive $\kappa$ for graphite dust, following Fox et al. (2010; their Figure\,4). For simplicity (and given the limited number of data points), we assume optically thin dust emitting at a single equilibrium temperature (e.g., Hildebrand 1983).  We calculated the dust temperature and mass up to day 815; later epochs are excluded owing to low S/N. Table\,4 summarizes the best-fitting parameters and limits. Our values are consistent with the results of Tinyanont et al. (2016), who performed similar calculations up to day 438.  Including the longer range of dates presented here, there is only marginal indication of a decrease in dust temperature and mass, considering the uncertainties. Overall, the ranges of mass and temperature values are broadly consistent with the current sample of SNe\,II (Tinyanont et al. 2016). \\

  \begin{table}
      \caption{Dust Emission Model Parameters.}
\renewcommand{\arraystretch}{1.7}
\centering
  \begin{tabular}{@{}lcc@{}} 
  \hline
 Epoch (day)   & Temp. (K)  & Mass ($10^{-3}\,{\rm M}_{\odot}$)  \\                            
\hline
\hline
 237.6  &$474^{+91}_{-65} $ &$8.7\pm6.2$ \\
 260.9  & $465^{+89}_{-62}$ &  $7.8\pm5.6$ \\
 439.2  & $357^{+51}_{-42} $ & $6.9\pm4.9$\\
 467.9  &  $340^{+90}_{-70} $ & $7.5\pm6.6$ \\
  472.7  &   $378^{+140}_{-105}$ & $2.5\pm2.3$\\
 815.6  &  $376^{+72}_{-221}$	 & $<0.16$ \\
\hline
\end{tabular}
\label{table:spitzer_model}
\vspace{3mm}
\end{table}

\section{Discussion}
\subsection{Origins of Continuum and Line Polarization}
SN\,2013ej exhibits the strongest and most persistent continuum and line polarization ever reported for a SN~II-P or II-L during the recombination phase, which indicates substantial asphericity in the explosion and/or circumstellar environment.  Modeling in \S2.5 shows that the general features of strong continuum polarization and prominent modulations across line features can be mutually generated by electron scattering in an oblate ellipsoidal envelope, viewed nearly edge-on, or dust scattering in an asymmetric distribution of CSM. We suspect that both mechanisms are potentially involved in the strong and peculiar spectropolarimetric evolution of SN\,2013ej. The relative contributions of each scattering mode with time are difficult to ascertain. However, the fact that the dust-scattering model does not fully reproduce the substantial depolarization of the emission lines during the recombination phase (for example, H$\alpha$ and Ca\,{\sc ii} on day 73), nor the narrowness and high radial velocity of the strongly polarized H$\alpha$ feature, suggests that electron scattering is probably the dominant mode of polarization for most of the recombination phase. 

By day 107, however, when recombination of the SN envelope is near complete, the increasing polarization level at the wavelengths of the emission peaks suggests that dust scattering becomes more influential as the optical depth to electron scattering drops. It is therefore interesting that on day 107, when the contribution of dust scattering might be at its maximum, the position angle remains within $\sim6^{\circ}$ of the average value during the earlier recombination phase. The position angle also never deviates more than $\sim13^{\circ}$ for the continuum at any time during the evolution of the SN. This implies that, if the spectropolarimetric evolution is indeed an evolving mix of dust scattering (CSM) and electron scattering (aspherical photosphere), then the two scattering geometries are probably similar. Below, we consider several physical scenarios that can potentially explain the spectropolarimetric data.

\subsubsection{Polarization via Aspherical CSM Interaction}
\begin{figure}
\includegraphics[width=3.3in]{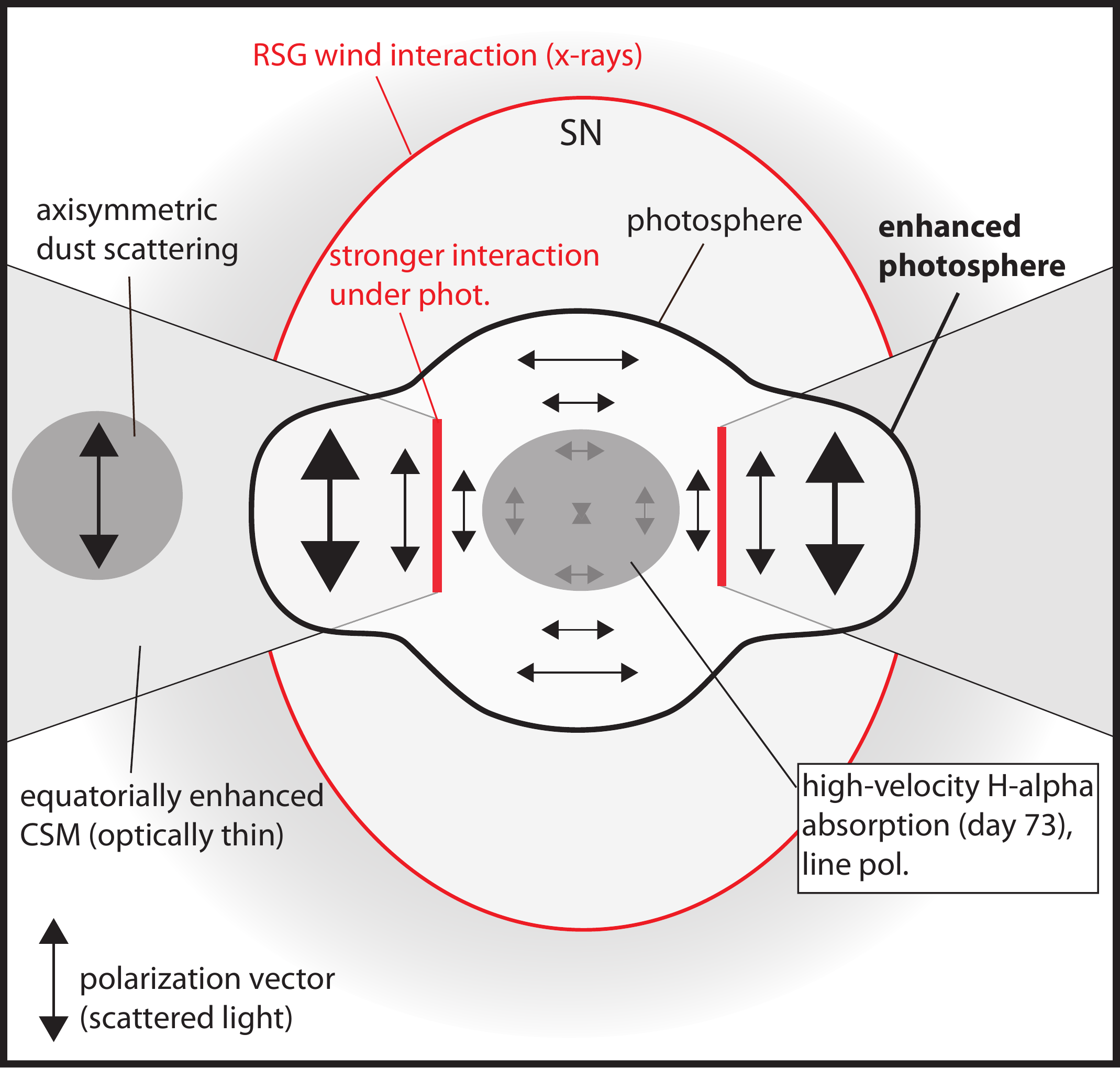}
\caption{Illustration of our proposed scenario for producing an oblate ellipsoidal photosphere via CSM interaction (observer's view). Geometric features are exaggerated for clarity, and not to scale.}
\label{fig:cartoon}
\vspace{3mm}
\end{figure}

We suspect that the unusual polarization characteristics of SN\,2013ej are the result of CSM interaction, which we already know was occurring during the recombination phase, based on the persistent X-ray emission (Chakraborti et al. 2016). We also know that the CSM interaction was geometrically asymmetric based on the multi-peaked morphology of broad H$\alpha$ emission at late times (see Figure\,\ref{fig:ha_velpol}). The aspherical component of the scattering geometry could thus be governed by the aspherical CSM distribution. Specifically, as the SN ejecta expand into the CSM, the added heat from interaction could result in a radially extended electron-scattering photosphere along directions where the CSM is densest, increasing the Thomson optical depth and producing excess continuum luminosity. For CSM that is in an equatorially enhanced distribution, perhaps stemming from mass loss in a binary system (e.g., common-envelope evolution and/or non-conservative mass transfer), the photosphere of the outflow could develop a bulged equator as the ejecta photosphere overtakes and envelops the region of intense interaction, where the shock slows. This sort of buried interaction was proposed to explain the temporary diminishing of narrow-line emission from the peculiar interacting SN PTF11iqb (Smith et al. 2015), as well as its relatively high luminosity, relative to normal SNe II-P, and fast decline rate during the plateau phase. An equatorially enhanced photosphere will result in a strong polarization signal, if viewed at a suitable inclination angle; we suggest this to be the case for SN\,2013ej. An illustration of this potential scenario is presented in Figure\,\ref{fig:cartoon}. Note, the detection of persistent X-rays by Chakraborti et al. (2016) implies CSM interaction outside the photosphere, with the RSG wind; the embedded equatorially enhanced interaction that potentially creates the ellipsoidal photosphere is a separate component. At late times, emission from the reverse shock-excited ejecta will be enhanced along the same equatorial direction, potentially creating the multi-peaked H$\alpha$ morphology we observe at $>800$ days. 

Since it is the CSM distribution that fosters the aspherical component of the photosphere, any dust in the outer reaches of the same distribution of CSM could result in an added source of scattering and polarization in a geometry that is axisymmetric with the photospheric scattering. In the case of an ellipsoidal photosphere from an equatorial distribution of dusty CSM, the two sets of Stokes vectors from dust scattering and electron scattering could constructively interfere, resulting in persistently strong polarization with little drift in position angle as their relative contributions change with time. Meanwhile, the SN photosphere along the more weakly interacting poles will recombine at a faster rate, temporarily amplifying the effect of the equatorial component of the photosphere. This could produce the rise in polarization strength and a small change in position angle shown in Figure\,\ref{fig:pt_lc}. As electron scattering diminishes, dust particles farther out in the same equatorially enhanced distribution of CSM could continue scattering and polarizing the SN light at a similar position angle.

The CSM-interaction scenario above could also explain the strong polarized H$\alpha$ feature and its high degree of axisymmetry with respect to the continuum. Indeed, the fact that its high radial velocity matches that of the outermost ejecta (see Figure\,\ref{fig:ha_velpol}) implies that the region of aspherical interaction at late times and the asymmetric H$\alpha$ absorption that is creating the polarized feature during the earlier recombination phase have a common origin in the outermost ejecta. As noted in \S2.4.1, the rapid development of the narrow component of the polarized H$\alpha$ feature coincided with the sudden appearance of the high-velocity ``notch" in the absorption component of the flux line profile (see Figure\,\ref{fig:linpol}, upper-left panel). Bose et al. (2015) previously interpreted the morphology of this absorption feature as the result of a separate high-velocity shell of H\,{\sc i} excited by CSM interaction. Similar signs of weak CSM interaction have been noted in other SNe II-P, such as SN\,2004dj (Chugai et al. 2007), SN\,2009bw (Inserra et al. 2012), and SN\,2012aw (Bose et al. 2013). 

The process by which CSM interaction influences the morphology of the H$\alpha$ absorption profile was investigated by Chugai et al. (2007); those authors showed that, upon encountering CSM, outer SN ejecta that would otherwise be optically thin are excited by X-rays from the forward and reverse shocks. This effect sustains high optical depth for H$\alpha$ in the outer high-velocity ejecta for a longer period of time than in the absence of CSM interaction, resulting in an observable feature in the blue shoulder of the absorption profile (e.g., a ``notch" or multiple minima). Since this added absorption occurs in the outermost ejecta, from the observer's perspective the absorption will obscure only the weakly polarized forward-scattered light from the central region of the photosphere, thus amplifying the effect of the highly polarized light scattered from the limbs of the photosphere (near $90^{\circ}$ scattering angle), producing the strong and narrow polarization peak at a fast radial velocity and with a position angle that is axisymmetric with the continuum. Radial variations in CSM density could thus lead to a rapid evolution of the blue side of the absorption profile, which, in turn, will result in a rapidly evolving polarized line feature. 

It is important to note that rapid changes in the polarized line spectrum can also occur in the dust-scattering (echo) scenario, since at any given time the total signal is a mix of direct SN light and highly polarized dust-scattered light from an earlier phase. Thus, if the SN suddenly develops enhanced line absorption, whether from enhanced CSM interaction or something else, the direct SN light will have a decreased contribution to the total flux. Since the more highly polarized echo spectrum has yet to reflect the additional absorption owing to larger light-travel time, there will be a temporary increase in the fraction of polarized light at the wavelength of the absorption enhancement, and a polarization feature will emerge and/or strengthen. We note that in the case of dust scattering, substantial deviations in position angle over line features are not expected. 

In each of the scenarios postulated above, it is the CSM and interaction that is responsible for the continuum polarization and the spectral changes that lead to the rapid development of enhanced line polarization. We note that similarly strong continuum and line polarization was observed in the Type IIb SN\,1993J, which also showed signs of asymmetric CSM interaction during its recombination phase and at late times (see Figure\,\ref{fig:latespec_compare}). This SN was also modeled as a nearly edge-on oblate ellipsoidal photosphere (H\"{o}flich et al. 1996), with potential contribution from circumstellar dust (Tran et al. 1997). It appears that SN\,2013ej and SN\,1993J could have exploded into similar asymmetric circumstellar environments of tenuous, dusty CSM. 

CSM interaction also might also explain the unusually early appearance of polarized Si\,{\sc ii} in SN\,2013ej from day 11. Indeed, the presence of this feature, first noted by Leonard et al. (2013), indicates peculiar ionization characteristics for a young SN~II, for which we suspect enhanced line optical depth from CSM interaction is potentially responsible. The lack of substantial changes in $\theta$ across this feature (see Figure\,\ref{fig:p_theta}), similar to the polarized H$\alpha$ feature, is also consistent with a high degree of axisymmetry.

\subsubsection{Polarization via Aspherical Explosion Geometry}
We must also consider the possibility that the electron-scattering mode of polarization is the result of an intrinsically aspherical explosion geometry, irrespective of CSM influence. In this case, the early onset of polarization would imply that the electron-scattering photosphere was aspherical while in the hydrogen envelope. In previous spectropolarimetric studies of normal  SNe~II-P, strong continuum polarization has only been observed to develop at relatively late times, near the onset of the nebular phase, after the photosphere has moved into the helium layer. Asphericity that appears only in deep layers has been attributed to an asymmetric distribution of radioactive material from the core (Wang et al. 2003; Leonard et al. 2006; Chornock et al. 2010). The resulting asymmetric distribution of heating produces an aspherical photosphere. Since radioactive asphericities are core-born, they are not expected to be observable at earlier phases when the photosphere is far out in the thick hydrogen envelope, unless such structures can penetrate the envelope rapidly --- for example, in the form of energetic jets of $^{56}$Ni. Simulations by Couch et al. (2009) showed that non-relativistic radioactive jets that have a high ratio of kinetic to thermal energy move ballistically, and can quickly punch through the thick RSG envelope and produce an aspherical SN photosphere at the earliest phases. We thus speculate that the polarization of SN\,2013ej could be the signature of such a jet-driven bipolar explosion. It is noteworthy in this regard that a bipolar distribution of $^{56}$Ni was invoked to explain the nebular-phase line profile of H$\alpha$ in SN\,2013ej (Bose et al. 2015; Yuan et al. 2016). The spectropolarimetry of a bipolar explosion, which could be approximated by a prolate ellipsoid, might appear similar to that of the oblate ellipsoidal configuration that we modeled in \S2.5.2. It is also plausible that such a bipolar explosion geometry results in the multi-peaked morphology of the broad H$\alpha$ emission that we observed at $>800$ days. The interpretation of a ballistic jet is also interesting in that the models from Couch et al. (2009) have shown that some of the the radioactive material overtakes the hydrogen envelope within the first week. We thus question whether the early appearance of Si\,{\sc ii} line polarization could be an additional signature of high-velocity $^{56}$Ni from a jet-driven explosion. 

\subsection{The Source of Warm Dust: Ejecta or CSM?}
The warm dust responsible for the mid-IR emission could involve multiple and physically distinct components. The light curve of SN\,2013ej and the inferred dust temperature and mass are broadly consistent with the properties of most SNe~II (Tinyanont et al. 2016), including several objects that exhibit evidence for interaction with tenuous CSM. Pre-existing dust in the circumstellar environment can be heated by the SN radiation and/or swept up and heated by the interaction shock, generating a source of mid-IR emission. Interestingly, our best-matching dust-scattering model for the spectropolarimetric data on SN\,2013ej involved a dust component located at a distance of $7.8\times10^{16}$\,km from the SN. Assuming a speed of 12,000\,km\,s$^{-1}$ for the fastest ejecta, this material would encounter the same light-scattering CSM dust at $\sim 800$ days, which is near the time we see the mid-IR emission possibly beginning to rise (see Figure\,\ref{fig:late_midir_lc}). In the relatable cases of SN\,2004et and SN\,2007od, both which exhibited late-time interaction-powered spectra similar to that of SN\,2013ej (see \S3), their thermal excesses were interpreted as a result of dust formation within a dense post-shock shell that developed in response to the CSM interaction (Kotak et al. 2009; Andrews et al. 2010; Inserra et al. 2011). This interpretation is therefore appealing for SN\,2013ej as well, not only because of its spectroscopic and SED similarities with SN\,2004et and SN\,2007od, but because of the possible signs of scattering by pre-existing circumstellar dust indicated by the spectropolarimetry in \S2.5.1. Interestingly, Leonard et al. (2009) noted that SN\,2004et also exhibited substantial continuum polarization $\sim3$ weeks into the photospheric phase, while preliminary spectropolarimetry of SN\,2007od also showed persistent continuum polarization at the level of $\sim0.8$\%--1.0\% throughout its photospheric phase (R. Chornock, private communication). All of these similarities seem to promote a physical connection between polarization, CSM interaction, and circumstellar dust.  

It is not clear, however, if the dust producing the mid-IR emission from SN\,2013ej is the same dust responsible for the red-side flux suppression of the broad H$\alpha$ profile on days 807 and 897. Indeed, this suggests that the obscuring dust is interior to the reverse shock, preferentially obscuring the receding ejecta on the far side of the outflow. If the obscuring dust were in a post-shock shell on the outside of the reverse shock, then the entire H$\alpha$ profile should be evenly obscured. A clumpy distribution of dust in the post-shock medium, as opposed to a continuous shell, might provide a solution, as some blueshifted emission from the near side of the reverse shock could potentially avoid obscuration, thus resulting in the asymmetric line profiles we observe. Such a clumpy distribution of dusty material was suggested in the relatable case of SN\,2007od (Andrews et al. 2010; Inserra et al. 2011).

Another possible source of thermal dust emission is the metal-rich SN ejecta, interior to the reverse shock. Such a configuration naturally results in a suppressed red side for the H$\alpha$ line profile from shock-excited ejecta. This hypothesis might also be more consistent with the $\sim10^{-3}\,{\rm M}_{\odot}$ dust masses inferred from our calculations in \S4.3.1 (Eq.~1), since such a pre-existing CSM dust mass would require a total CSM much more massive than expected around a normal RSG, and more on par with strongly interacting SNe~IIn (e.g., see Fox et al. 2011) --- unless the CSM had a different origin than an RSG wind (e.g., equatorial mass loss via binary interactions). Of course, it is also possible that the dust producing the mid-IR emission is not the same dust responsible for the uneven obscuration of the H$\alpha$ emission. Indeed, as stated previously, multiple components of dust within the ejecta, CSM, and post-shock medium could contribute to the mid-IR evolution.

\section{Summary and Concluding Remarks}
We have presented spectropolarimetry, deep late-time spectroscopy, and space-based UV--IR photometry of the weakly interacting SN\,2013ej. From day 11 through 107, the SN exhibits the strongest and most persistent continuum polarization ever reported for a SN II-P or II-L, and also unusually prominent and narrowly peaked line polarization for this type of SN, particularly after $\sim2$ months. Preliminary modeling in \S2.5.1 has shown that the continuum polarization and broad modulations across the line features can be mutually generated by an ellipsoidal photosphere, and by optical scattering off dust particles in an asymmetric distribution of CSM. The spectropolarimetric evolution of SN\,2013ej could be governed by a complex time-dependent combination of Stokes vectors from both of these scattering processes. However, electron scattering in an ellipsoidal photosphere is likely to be the dominant source of polarization, particularly on day 73, when nearly full depolarization of the H$\alpha$ and Ca\,{\sc ii} emission-line peaks is observed.  The dust-scattering mode appears to become more influential on our last epoch at the end of the recombination phase (day 107), based on the increased amount of residual polarization for the H$\alpha$ emission peak, and on the reasonable match of our model. The continued increase in fractional polarization indicated by the late-time imaging polarimetry from Kumar et al. (2016) could indicate that scattering by circumstellar dust continues beyond day 130, or that the asphericity in the inner electron-scattering environment became even more pronounced. 

Deep optical spectroscopy past 800 days reveals persistent and geometrically asymmetric CSM interaction into late phases, in the form of broad multi-peaked emission-line profiles and a blue pseudocontinuum. The broad spectral features are indicative of SN ejecta energized by the passage of the reverse shock, and exhibit similarities with other well-studied core-collapse SNe that have exhibited evidence for persistent interaction, substantial polarization, and late-time thermal excess from dust, including SN\,1993J (Matheson et al. 2000; H\"oflich et al. 1996; Tran et al. 1997), SN\,2004et (Kotak et al. 2009; polarization noted by Leonard et al. 2009), and SN\,2007od (Andrews et al. 2010; polarization information from R. Chornock, private communication). The late time UV-optical photometry of SN\,2013ej from HST, out to day 1169, which is much brighter than what is expected from radioactive decay, is also indicative of ongoing CSM interaction. The CSM responsible for this late-time interaction was probably created at least $\sim3000$\,yr before core collapse. The hint of triple-peaked line morphology is interesting in the context of the spectropolarimetry, as it appears to be consistent with an equatorially enhanced distribution of CSM, perhaps stemming from mass loss in a binary system. 

We currently lean toward a scenario in which interaction with an equatorially enhanced distribution of CSM, viewed nearly edge-on, is responsible for the photospheric asphericity, the resulting polarization, and (perhaps) for providing some additional luminosity early in the recombination phase that may have influenced the relatively rapid decline rate. Dust in the same distribution of CSM could provide an additional source of scattering that is nearly axisymmetric with the interaction-influenced photosphere, constructively adding to the net polarization.  The late-time mid-IR emission detected with \textit{Spitzer} out to $\sim1000$ days could involve the same circumstellar dust that potentially contributes to the optical scattering at earlier phases, radiatively heated by the propagating interaction shock. This scenario is reminiscent of interpretations previously suggested to explain the spectropolarimetry of the Type IIb SN\,1993J (H\"oflich et al. 1996), and of more strongly interacting SNe\,IIn, such as SN\,2009ip (Mauerhan et al. 2014), SN\,1998S (Leonard et al. 2000; Wang et al 2001), and SN\,1997eg (Hoffman et al. 2008). It is noteworthy that, if viewed at a suitable inclination angle, weakly interacting SNe can potentially produce continuum and line polarization levels comparable to those of SNe~IIn, even if the CSM densities and shock strengths are not high enough to generate the strong narrow emission lines that characterize the flux spectra of SNe~IIn. Future studies involving a larger sample of objects for which there is high-quality spectropolarimetry, deep late-time spectroscopy, and mid-IR photometry will further elucidate the physical connections between CSM interaction, asphericity/polarization, light-curve evolution, and thermal dust emission.

We also considered the alternative scenario that the polarization is produced by an intrinsically aspherical explosion, perhaps the result of a ballistic jet that drives a bipolar outflow. If this interpretation is correct, it would be remarkable that such a ballistic jet could result from the collapse of a rather modest 9.5--15.5\,M$_{\odot}$ RSG, which is the mass estimated for the directly identified progenitor of SN\,2013ej (Fraser et al. 2014).

We note that, although the physical interpretations we have considered are plausible and guided by theoretical expectations, none offer unique solution to the data on SN\,2013ej. Models involving more detailed treatment of complex scattering geometries and their combined spectropolarimetric effects are required, and we look forward to their construction in the future.

Finally, the late-time UV--optical photometry from {\it HST} shows that the interaction-powered light curve faded (temporarily, at least) well below the level of the RSG progenitor, confirming its association with the explosion (Fraser et al. 2013; Van Dyk et al. 2013). 

\section*{Acknowledgements}
\scriptsize 
J.C.M. would like to thank Nathan Smith for informative scientific discussion and commentary on this manuscript. Support for {\it HST} programs GO-14116 and GO-14668 was provided by NASA through grants from the Space Telescope Science Institute, which is operated by the Association of Universities for Research in Astronomy, Inc., under NASA contract NAS5-26555. We are grateful to the staffs at Lick and Keck Observatories for their excellent assistance. This work is based in part on observations from the Low Resolution Imaging Spectrometer at the Keck-1 telescope. The W. M. Keck Observatory is operated as a scientific partnership among the California Institute of Technology, the University of California, and NASA; it was made possible by the generous financial support of the W. M. Keck Foundation. We extend special thanks to those of Hawaiian ancestry on whose sacred mountain we are privileged to be guests. A.V.F.'s group at U.C. Berkeley is supported by NSF grant AST-1211916, Gary \& Cynthia Bengier, the Richard \& Rhoda Goldman Fund, the Christopher R. Redlich Fund, and the TABASGO Foundation.  Research at Lick Observatory is partially supported by a generous gift from Google. This work is based in part on observations and in part on archival data obtained with the {\it Spitzer Space Telescope}, which is operated by the Jet Propulsion Laboratory, California Institute of Technology, under a contract with NASA; support was provided by NASA through an award issued by JPL/Caltech.

\end{document}